\documentclass[11pt,twoside]{article}
\usepackage{cozumel2005}
\usepackage{epsf}
\usepackage{psfig}
\usepackage{lscape}
\begin{document}
\title{Exotic populations in Galactic Globular Clusters}    %%% Fill in title
\author{Francesco R. Ferraro}   %%% Fill in author names
\affil{Dipartimento di Astronomia, Universit\`a di Bologna, Via
  Ranzani 1, 40127 Bologna (ITALY)}    %%% Fill in author affiliations

\begin{abstract} %%% Abstract to run on from here.
Recent high-resolution observations of the central
region of Galactic globular clusters have shown the presence
of a large variety of exotic stellar objects whose formation
 and evolution may be strongly affected by dynamical
 interactions.
In this paper I  review the main properties of two
classes of exotic objects:
 the so-called Blue Stragglers stars and the recently
 identified optical companions to Millisecond pulsar.
 Both these  class of objects are invaluable tools to
 investigate the binary evolution in very dense environments
 and   are powerful tracers of the dynamical history
 of the parent cluster. 
\end{abstract}

%%% MAIN BODY OF TEXT GOES HERE. CONSULT  manual_cozumel2005.tex 
%%% SECTIONS 2.3-2.6 FOR HELP WITH EQUATIONS, FIGURES,
%%% AND TABLES.
 
\section{Introduction}

Ultra-dense cores of Galactic Globular Clusters (GCs)  are
very efficient ``furnaces'' for generating exotic objects, 
such as low-mass X-ray binaries, cataclysmic variables,
millisecond pulsars (MSPs), blue stragglers (BSS), etc.
Most of these objects are thought to result from the
evolution of various kinds of binary systems originated
and/or hardened by stellar interactions. The nature and
even  the existence of binary by-products can be strongly 
affected by the cluster core dynamics, thus   serving as a
diagnostic  of the dynamical evolution of GCs.  This topic
has received strong impulse in the recent years and many
studies have been devoted to investigate the possible link
between the dynamical  evolution of clusters and the
evolution of their stellar population. 

In particular two
aspects can be investigated: (1) the environment effects
on canonical evolutionary sequences (as for example the
possible  effect of different environments on the  blue tail
extension of the Horizontal Branch - HB, see Fusi Pecci et
al 1992, Buonanno et al 1997); (2) the creation of
artificial sequences as BSS and other exotic objects (see
Bailyn 1995 and reference therein). 

In this paper I will
review the main properties of two anomalous
sequences in the color-magnitude diagram (CMD):

\begin{itemize}
\item the most  known {\it anomalous sequence}
in GCs: the so-called BSS sequence, indeed the very first
sequence of exotic objects discovered in the CMD of GCs;
\item the most recently discovered {\it anomalous sequence}:
the one defined by MSP companions.
\end{itemize}

\section{Blue Straggler Stars} 

BSS, first discovered by Sandage (1953) in M3,
  are commonly defined as   stars  
  brighter and bluer (hotter) than the main sequence (MS) turnoff
(TO), lying along an apparent extension of the MS, and thus
mimicking a rejuvenated stellar population. The existence of such a
population has been a puzzle for many years, and even now its
formation mechanism is not completely understood, yet. At present,
the leading explanations involve mass transfer between binary
companions,  the merger of a binary star system or the collision of
stars (whether or not in a binary system). Direct measurements
(Shara et al. 1997) 
and indirect evidence  show that BSS are
more massive than the normal MS stars, pointing again
towards collision or merger of stars. Thus, BSS represent the link
between classical stellar evolution and dynamical processes
(see Bailyn 1995). The
realization that BSS are the ideal diagnostic tool for a
quantitative evaluation of the   dynamical interaction effects
inside star clusters has led to a remarkable burst of searches and
systematic studies, using UV and optical broad-band photometry.

\subsection{The UV approach to the study of BSS} 
 
The observational and interpretative scenario of BSS has
significantly changed in the last 20 years. In fact, since
their discovery and for almost 40 years, BSS have been
detected only in the outer regions of GCs or in relatively
loose clusters, thus forming the idea that a low-density
environment is the {\it natural habitat} for BSS.
 Of course, it was an observational bias: starting from the early
 '90 high resolution studies allowed to properly image the
 central region of high density clusters (see the case of 
 NGC6397 by Auriere \& Ortolani 1990). Moreover, 
 with the advent of the Hubble Space Telescope ({\rm HST}) it became
possible for the first time to search dense cluster cores for
BSS. This was a really turning point in BSS studies since 
HST, thanks to its
unprecedented spatial resolution and  imaging/spectroscopic
capabilities in the UV, has given a new impulse to the
study of  BSS (see Paresce et al 1992, Paresce \& Ferraro
(1993), Guhatahakurtha et al 1994, etc).

\begin{figure}[!ht]
  \plotone{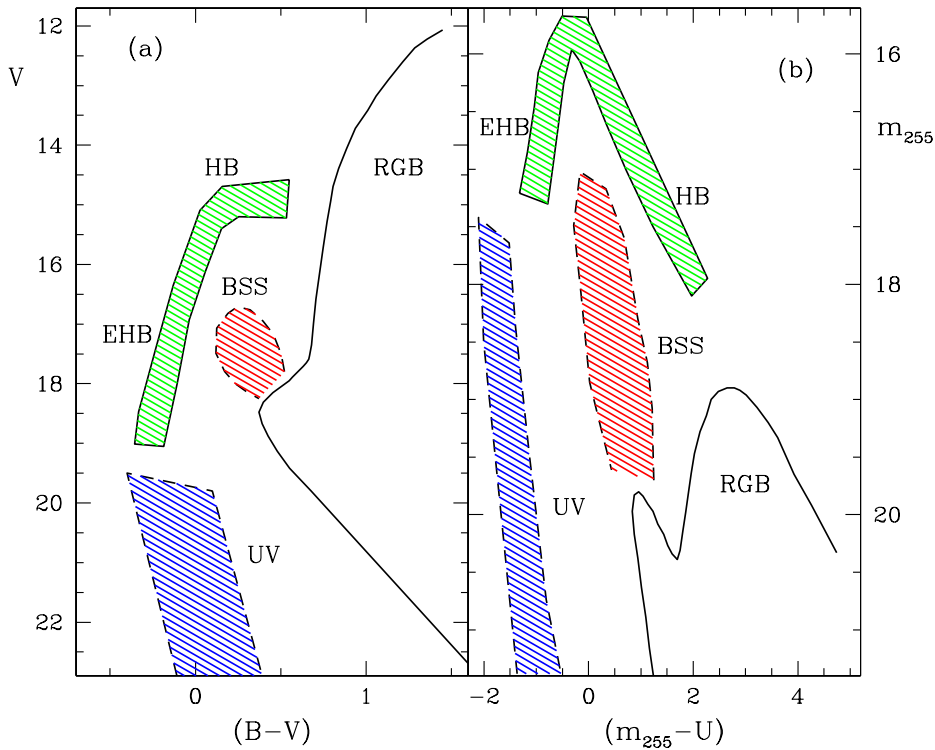}
  \caption{ The main sequences of a GC CMD
in the (V,B-V) and ($m_{255},m_{255}-U$)-planes, respectively.}
 \end{figure}

Based on  these observations, the first catalogs of BSS
have been published
  (Fusi Pecci et al 1992;
Sarajedini 1992;  Ferraro, Bellazzini \& Fusi Pecci
1995, hereafter FFB95)
until the most recent collection of BSS which 
counts nearly 3000 candidates
(Piotto et al 2004). 
These works have significantly
contributed to form the nowadays, commonly accepted idea
that BSS are a normal  stellar
population in clusters, since they are present in all of
the properly observed GCs. However, according to Fusi
Pecci et al.  (1992) BSS in different environments could
have different origin. In particular, BSS in loose GCs
might be produced from coalescence of primordial binaries,
while in high density GCs (depending on
survival-destruction rates for primordial binaries) BSS
might arise mostly from stellar interactions, particularly
those which involve binaries. Thus, while the suggested
mechanisms for BSS formation could be at work in clusters
with different environments (FFB95; 
Ferraro et al. 1999),  there are evidence that
they could also act simultaneously within the same cluster
(as in the case of M3, see Ferraro et al. 1993 - hereafter
F93; Ferraro et
al. 1997 -  hereafter F97).  
Moreover, as shown by Ferraro et al. (2003 - hereafter F03),
both the BSS formation channels (primordial binary
coalescence and stellar interactions) seem to be equally
efficient in producing BSS in different environments, since
the two clusters that show the largest known BSS specific
frequency, i.e. NGC~288 (Bellazzini et al. 2002) and M~80
(Ferraro et al. 1999), represent two extreme cases of
central density concentration among the GCs ($Log
\rho_0=2.1$ and $5.8$, respectively).

However,  a major problem   in the systematic
study of the BSS still persists, especially in the central region of high
density clusters by using the classical  CMD even with HST.
In fact, the  CMD of an old stellar population (as a GC)
in the {\it classical} $(V,B-V)$ plane
is dominated  by
the cool stellar component, hence the observations and the
construction of complete sample of hot stars 
(as extreme blue HB, BSS, various by-products of binary 
system evolution etc.) is ``intrinsically''
difficult in this plane. Moreover, in visible CMDs the BSS
region could be severely affected by  photometric blends
which mimic BSS. 

In the ultraviolet (UV) plane, where
the sub-giants and red giant stars which cause BSS-like
blends are faint and the hot stellar populations are
relatively bright, problems are much
less severe, allowing to
obtain complete BSS samples  even in the
densest cluster core regions The advantage of studying BSS in
the mid-UV CMD is shown in Figure 1, where the shapes of the 
main evolutionary
sequences in the traditional  (V,B-V) plane ({\it panel
(a)}) and in the UV plane ({\it panel (b)}) are  compared.
As can be seen,
in the UV plane the main branches
display very different morphologies with respect to
those in  the  optical CMD
(i.e. $V,~ V-I$).    As can be seen, the red giant
branch (RGB) is very faint in the UV, while the HB,
excluding the hottest section, which bends downward because of the
increasing bolometric correction, appears diagonal. Since red giants
are faint in UV, the photometric blends, which mimic BSS in visible
CMDs, are less problematic. 
The BSS define a narrow, nearly vertical sequence spanning
$\sim 3$ mag in this plane, thus, a complete BSS
sample can be obtained even in the densest cores: indeed, the
($m_{255}$, $m_{255} - m_{336}$) plane is an ideal tool for selecting
BSS.

\begin{figure}[!ht]
  \plotone{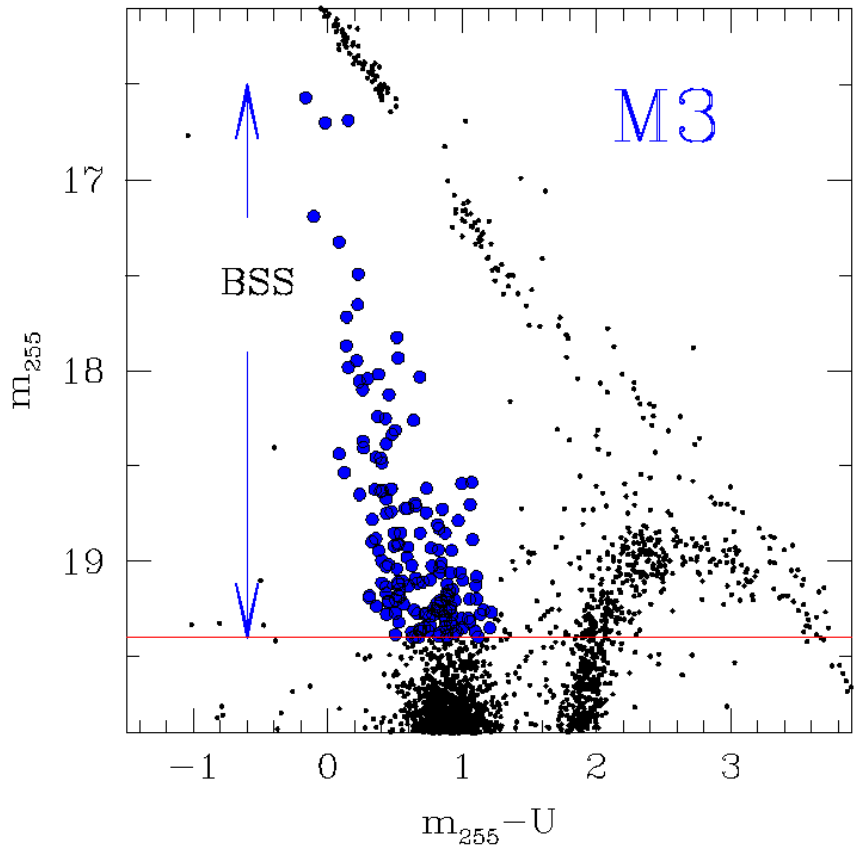}
  \caption{BSS in the UV: the case  of M3.
  The horizontal line at  $m_{255}=19.4$ is the 
  assumed limiting magnitude ($\sim 5
  \sigma$ above the TO level (from F97).}
\end{figure}

\subsection{M3: a new approach to  study
GC stellar populations}

M3 has played a fundamental role in the BSS history, since
it is the  GC where BSS have been identified for the first
time,  but also the first GC in which the BSS radial
distribution has been studied over the entire cluster
extension. In fact by combining UV HST observations in the
central  region of the cluster (F97, see Figure 2) and 
extensive wide field
ground-based observations (F93 -  Buonanno et
al 1994), F97 presented  the radial distribution of  BSS
over the entire radial extension ($r\sim 6 '$).

The UV search for BSS in the central region of M3 led to
the discovery of  a large population of  {\it centrally
segregated} BSS (see Figure 2),  contrary to the previous
claim by Bolte, Hesser \& Stetson (1993) who suggested a
possible  depletion of BSS in the central region of the 
cluster with respect to the external regions. 
As expected,
the BSS candidates occupy a narrow, nearly vertical,
sequence spanning $\sim 3$ mag in $m_{255}$. Two limits
(one in color, and one in magnitude) have been assumed to
properly select the BSS sample in the UV-CMD shown in
Figure~2.  The BSS sequence blends smoothly into the 
MS near the cluster TO. In order to select  `safe'
BSS, only stars brighter than $m_{255} \sim
19.4$ (0.3 mag brighter than the cluster TO) have been
considered.

Beside this, a much more
unexpected result was found from the analysis 
of BSS in M3. In fact, in order to 
extend  the BSS analysis to the whole radial extension of
the cluster and to  compare the HST sample
with the external (ground-based) one, F97 limited the
BSS analysis  to the brighter portion ($m_{255}<19$)
of  BSS population. The   radial distribution of the  
BSS candidates    was  compared to
that of a sample of RGB stars assumed as ``reference''
population.  The cumulative radial distributions of the
entire sample split into two sub-sets (at $r=150\arcsec$)
are reported in Figure 3: as can be seen  the BSS (solid
line)  are more centrally concentrated than RGB stars 
(dotted line) in the central regions (out to
r$<150\arcsec$), while are less concentrated in the outer
ones.

\begin{figure}[!ht]
  \plotone{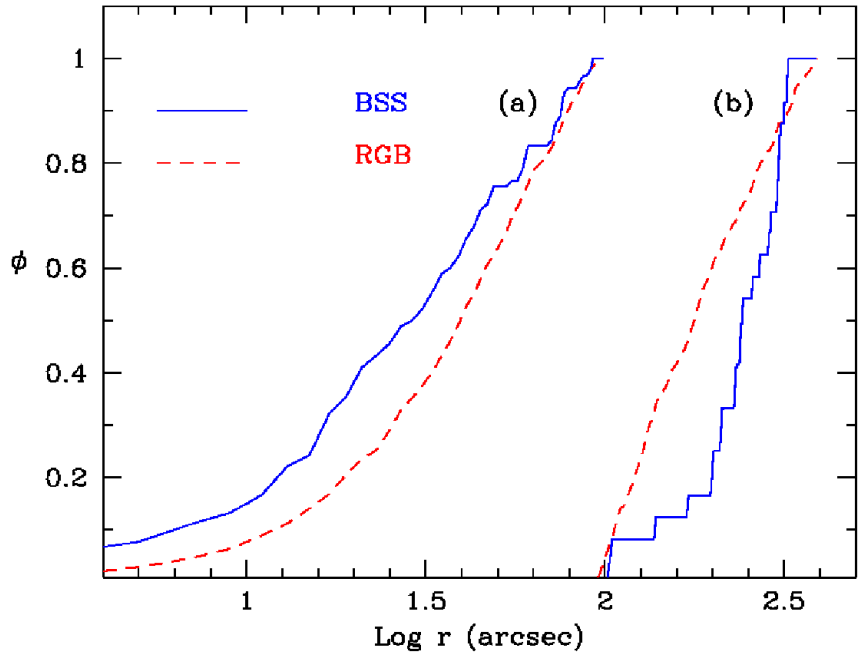}
  \caption{Cumulative distribution of bright BSS (solid
  lines) in M3
  with respect to the RGB (dashed lines), for two radial
  sub-samples: (a) $r<2'.5$ and (b) $2'.5<r<6'$ (from F97).}
\end{figure}

In order to further investigate  this surprising result,
F97  computed the doubly
normalized ratios for the BSS and the RGB stars, following the
definitions by F93:

\begin{displaymath}
R_{\rm BSS} = {{(N_{\rm BSS}/N_{\rm BSS}^{\rm tot})} \over 
{(L^{sample}/L_{tot}^{sample})}} 
\end{displaymath}

and

\begin{displaymath}
R_{\rm RGB} = {{(N_{\rm RGB}/N_{\rm RGB}^{\rm tot})} \over 
{(L^{\rm sample}/L_{\rm tot}^{\rm sample})}} 
\end{displaymath}

\noindent
respectively.

The surveyed cluster region has been divided in a number
of  concentric annuli and the numbers of BSS and RGB
counted  in each annulus has been normalized to  the
sampled luminosity accordingly to the above relations. The
{\it relative frequency} of BSS  is  compared
with that computed for the RGB ``reference'' stars  as a
function of the distance from the cluster center, as shown
 in Figure 4.  
As can be seen, the radial distribution of BSS is
clearly bimodal: it  reaches its maximum at the  center of
the cluster (showing no evidence of a BSS depletion in the
core); it has  a clear-cut dip in the intermediate region
(at $100\arcsec<r<200\arcsec$) and  a rising  trend in the
outer region (out to $r\sim360\arcsec$), as first noted  by
F93.

\begin{figure}[!ht]
\plotone{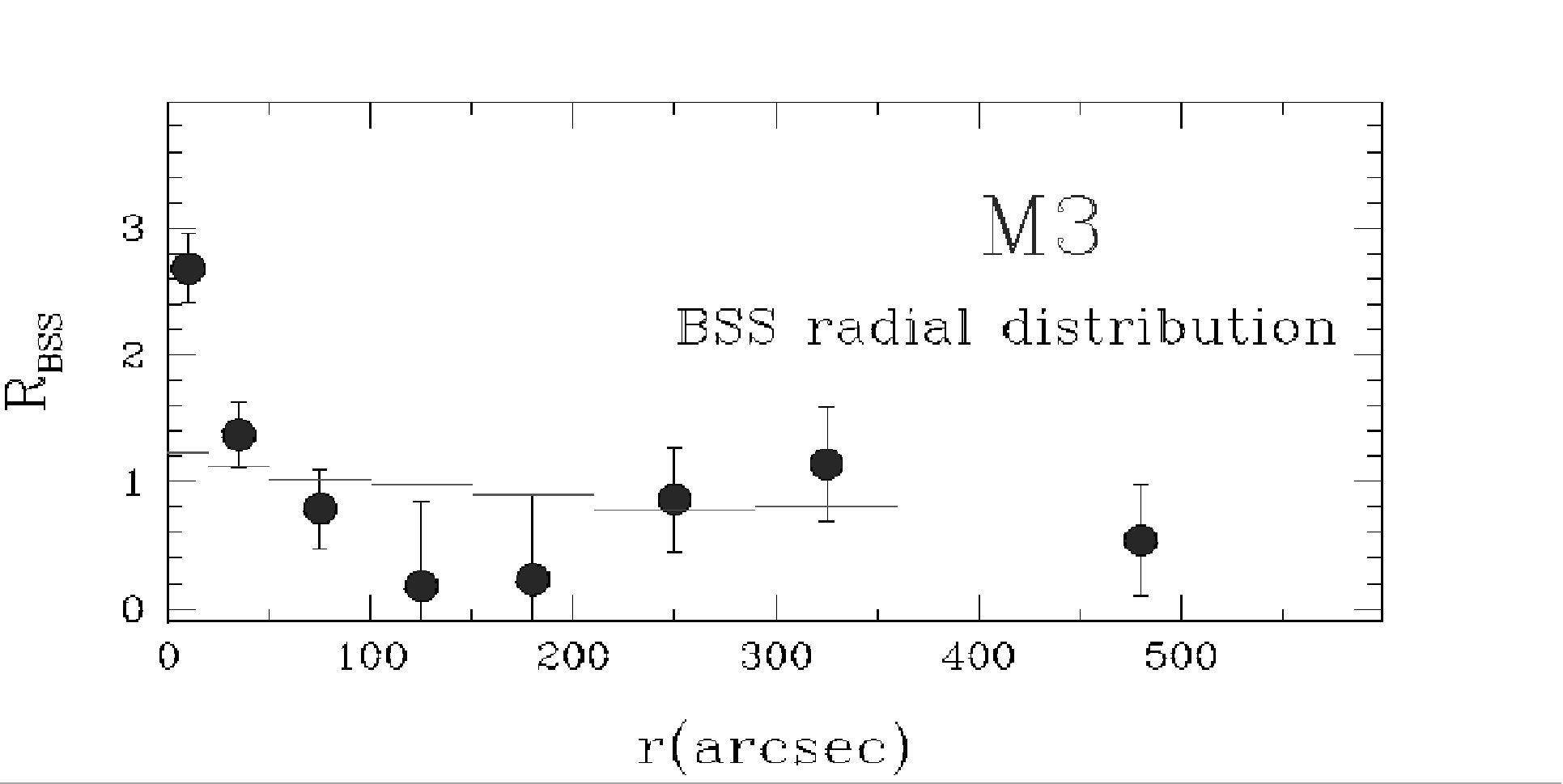}
\caption{ The relative frequency of BSS in M3 
is plotted as a function of the radial 
distance from the cluster center. The horizontal lines show the relative 
frequency of the RGB stars used as a comparison population.
For $r>6'$ only the relative frequency of BSS has been 
computed using the  Sandage (1953) candidates. (From F97)}
\end{figure}

\begin{figure}[!ht]
\plotone{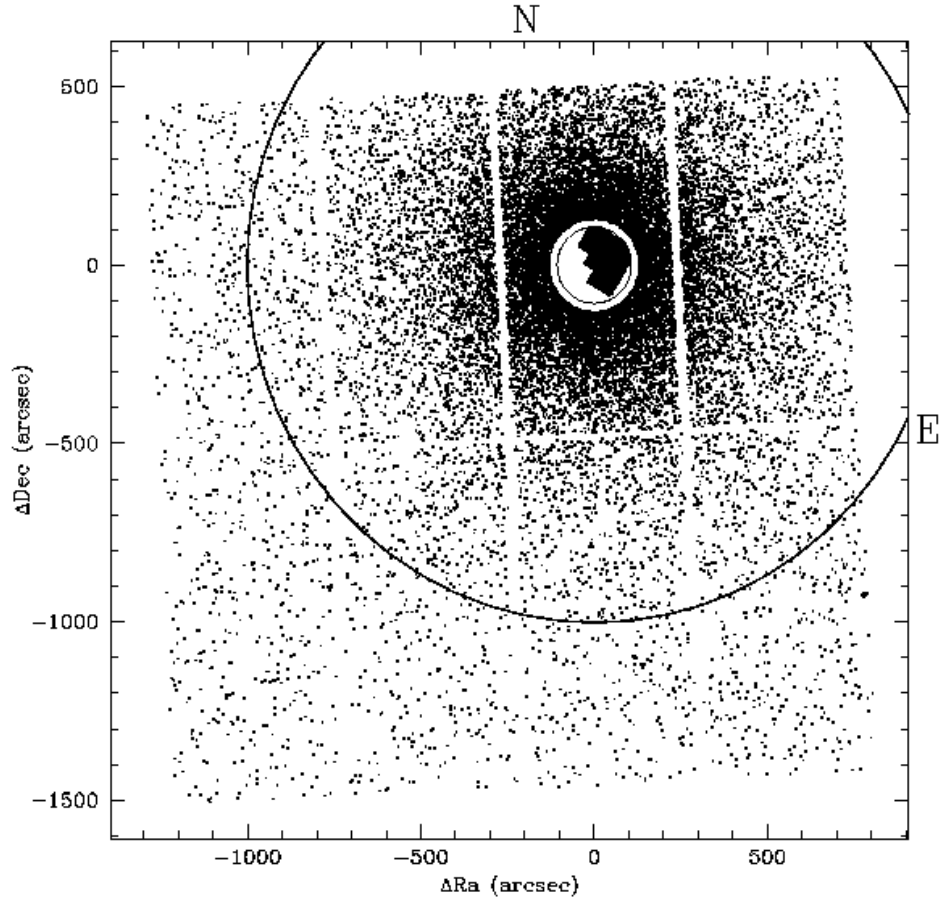}
\caption{The map shows the observational strategy adopted to
study the radial distribution of the BSS frequency in
clusters. Typically the central regions are imaged 
by UV-HST observations, while the external regions are
sampled by ground-based wide-field observations (using the
Wide Field Imager (WFI) at the 2.2m ESO telescope).
Here the case of NGC6752 is shown. 
 (From Sabbi et al 2004).}
\end{figure}

\begin{figure}[!ht]
\plotone{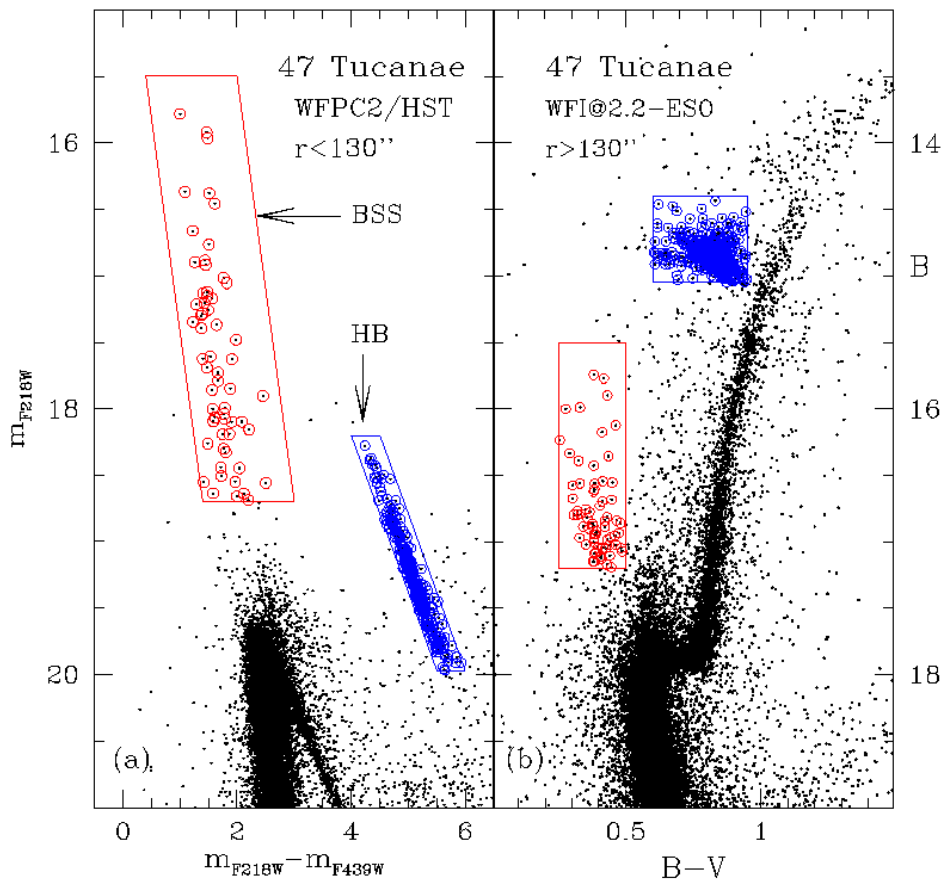}
\caption{{\it Panel (a):} ($m_{218}, m_{218}-m_{439}$) CMD
of the 
central region of 47 Tuc ($r<130''$) from
WFPC2/HST observations.
{\it Panel (b):} ($B,B-V$) CMD for the external part
 of 47 Tuc ($130''<r<1500''$) from ground-based
 observations. The two selection boxes for the BSS and the
 reference HB population are shown. (From Ferraro et al
 2001a and F04).}
\end{figure}

\begin{figure}[!ht]
\plotone{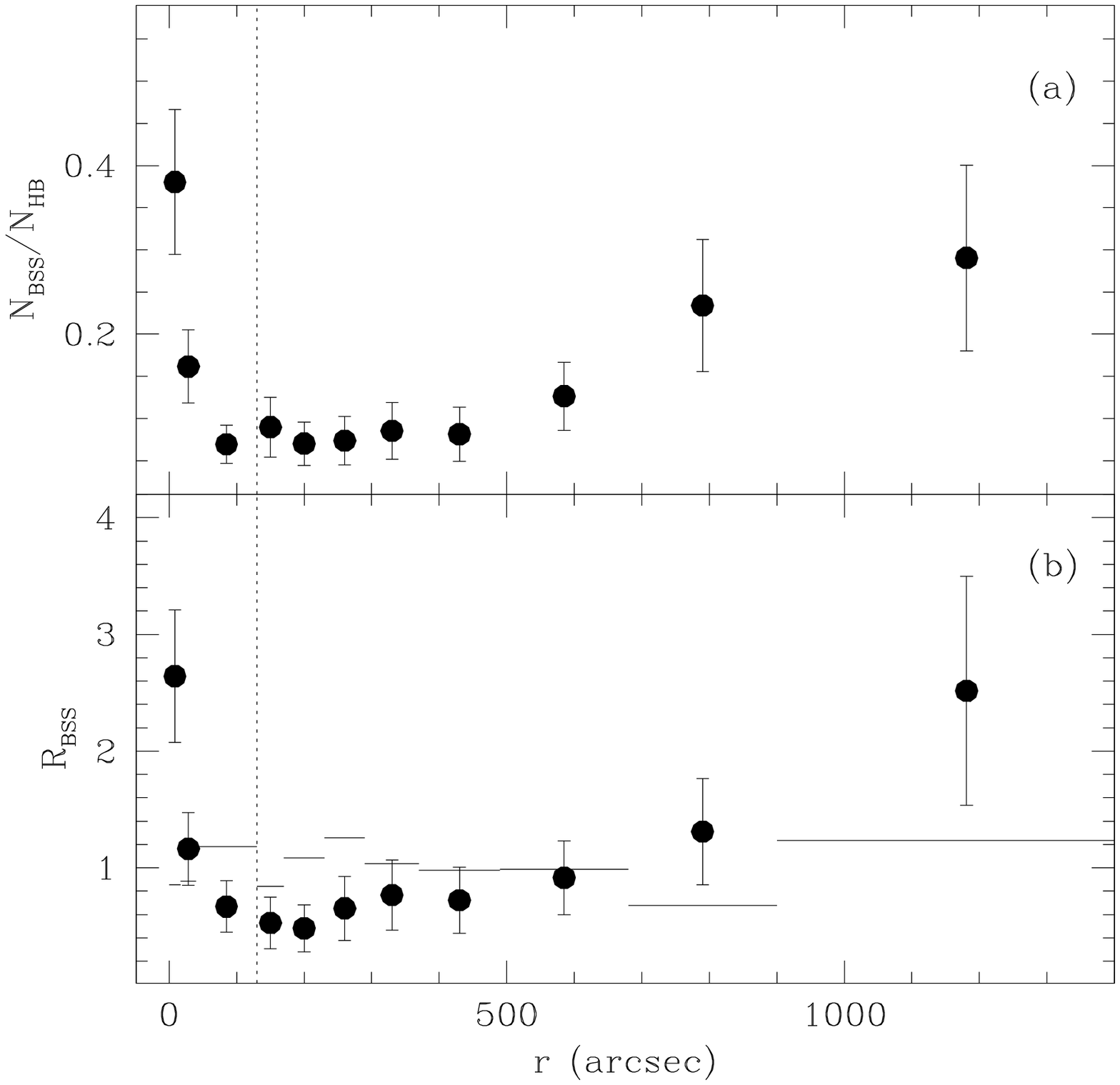}
\caption{{\it Panel (a):} The relative frequency
of BSS with respect to HB stars is plotted as a function
of the distance from the cluster center.
 {\it Panel (b):} The specific frequency of BSS  is
 plotted. The horizontal lines show the specific
 frequency for the HB reference population. 
 The vertical dotted line separates the cluster regions
observed   with HST from those observed from the
ground. (From F04).}
\end{figure}

\subsection{47 Tuc: another surprise!} 

While the bimodality detected in M3 was considered for
years to be {\it peculiar}, the most recent results demonstrated
that this is not the case.  In fact, in the last years 
the same observational strategy adopted by F97  in M3  has
been applied  to a number of clusters with the aim of
determining the BSS frequency over the entire cluster
extent.  
To do this  two data-set are generally combined: 

{\it (i) High resolution set---} consisting of a series of high-resolution
WFPC2-HST images  (typically in the UV) of the cluster
center. In this data set the planetary
camera (PC, which has the highest resolution $\sim 0\farcs{046}/{\rm
pixel}$) is roughly centered on the cluster center while the Wide
Field (WF) cameras (at lower resolution $\sim 0\farcs{1}/{\rm pixel}$)
sample the surrounding outer regions;

{\it (ii) Wide Field set---} consisting of a series 
of  multi-filter ($B$, $V$,
$I$) wide field images obtained by using the last generation
of wide field imagers (as for example 
the Wide Field Imager (WFI) mounted at the 2.2m ESO-MPI
telescope at ESO (La Silla)). 
The WFI is a mosaic of 8 CCD chips (each one 
with a field of view of $8'\times
16'$) giving a global  coverage of $33'\times 34'$. 
    
\begin{figure}[!ht]
\plotone{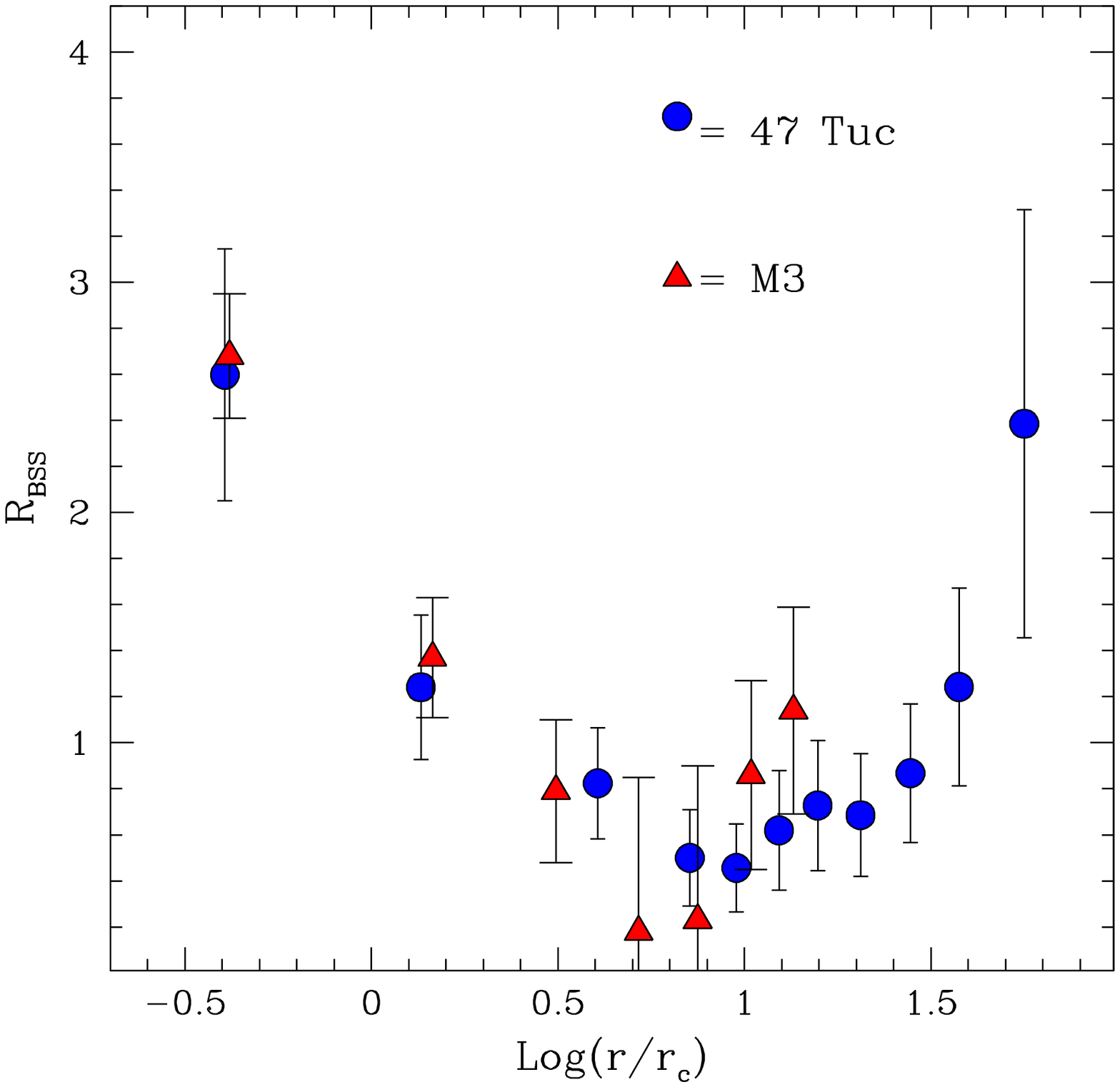}
\caption{ The specific frequency of BSS in 47 Tuc
(filled circles) is compared with that found in M3 (filled
triangles). The radial coordinate is expressed in unit of
core radii $r_c$.  (From F04).}
\end{figure}

Figure 5 shows the typical cluster coverage adopting this
observing strategy, also used  for 47 Tuc
(see Ferraro et al 2001a and Ferraro et al 2004a, hereafter F04).
Figure 6 shows the corresponding CMD. As can be seen a well
defined sequence of BSS has been obtained in both  
samples. To study the BSS radial distribution, 
F04 applied  the procedure described
in F93 (used also in F97) where the surveyed area has been divided
into a set of concentric annuli. 11 concentric annuli, each one
containing roughly $\sim 10$\% of the reference population) have been
defined in the case of 47 Tuc.
The BSS specific frequency has been computed in two different
ways: (1) the ratio $F_{\rm BSS}^{HB}=N_{\rm BSS}/N_{\rm HB}$
and (2)  the double-normalized ratio $R$ (see above)
 considering the fraction of luminosity
sampled in each annulus. 
 
Figure 7  shows the distribution of both ratios as a
function of the effective radius of each annulus. The distribution is
clearly bimodal, with the highest value in the innermost annulus
where 
the $F_{\rm BSS}^{HB}$ ratio reaches $\sim 0.4$, 
it
significantly decreases to less than 0.1 as $r$ increases and then
slowly rises  up to $\sim 0.3$ in the
outer region.   

\begin{figure}[!ht]
\plotone{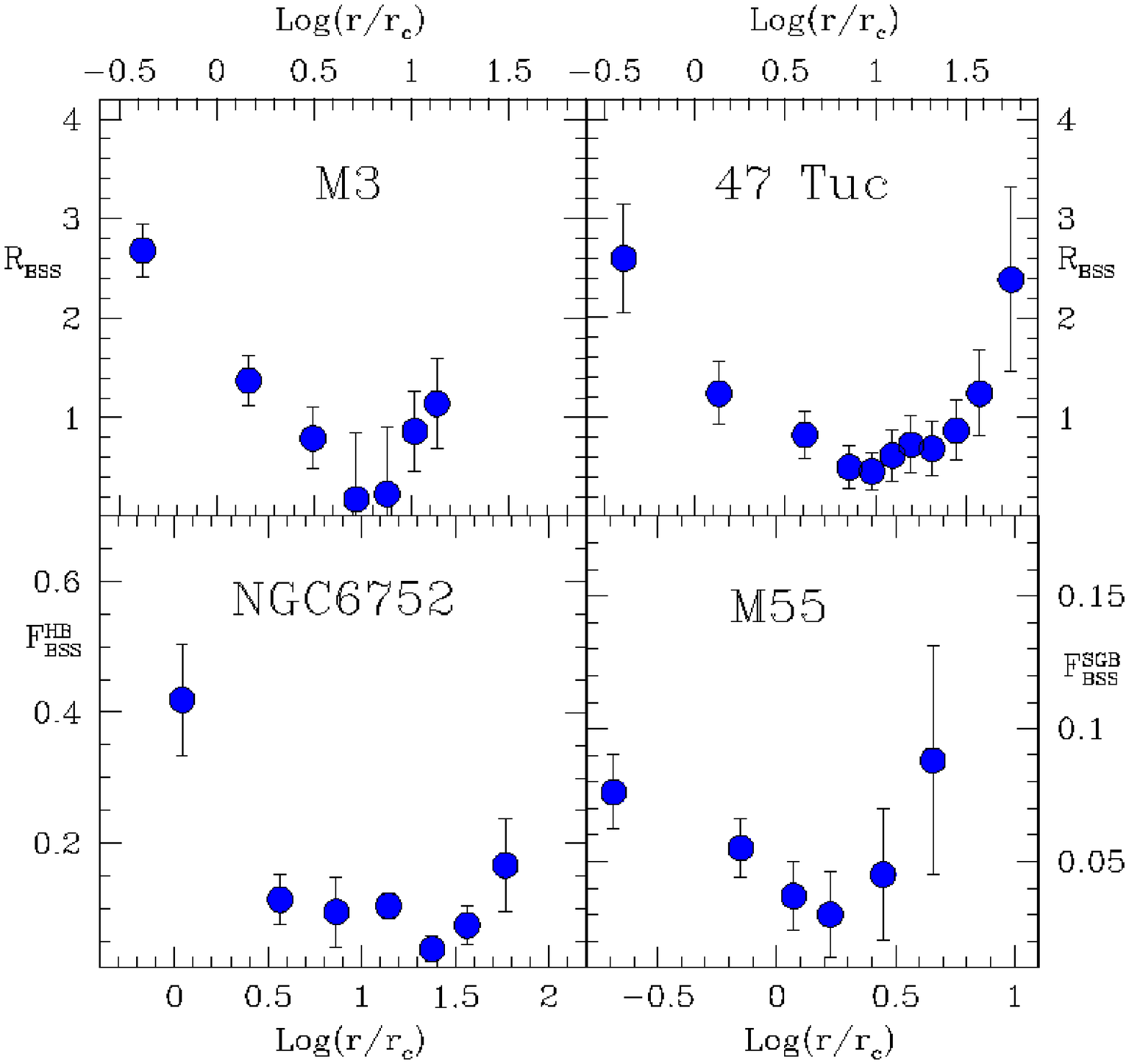}
\caption{ The bimodal distribution of BSS discovered
in GCs up to now.   
}
\end{figure}

This trend is fully confirmed by using the  
 {\it relative BSS frequency} $R_{BSS}$ (see F97 and above).
 The behavior of this ratio as a function of the
distance from the cluster center is shown in Figure
7 {\it (panel (b))} and compared with the corresponding one for
the HB ``reference'' stars. As can be seen, the HB specific
frequency remains essentially constant over the surveyed
area since the fraction of HB stars (as any post-main
sequence stage) in each annulus strictly depends on the
fraction of luminosity sampled in that annulus (see the
relation by Renzini \& Buzzoni  1986, eq 2 in F03).  
In contrast, the BSS specific frequency reaches its
maximum at the center of the cluster, then decreases to an
approximately constant value in the range
100\arcsec--500\arcsec\ from the cluster center and then
rises again. {\it The trend found in 47 Tuc    
closely resembles that discovered in M3 by F97}.

To further demonstrate the similarity with M3 and to study possible
differences we show the BSS specific frequencies for the two clusters on
the same figure (see Figure 8). The radial coordinate  is
given in units of the core radius $r_c$,
adopting   $r_c=21\arcsec$ and
$r_c=24\arcsec$ for 47 Tuc and M3, respectively (see F04 and
F03). A few major characteristics about Figure 8
are worth noticing: (1) the central values are similar; (2)
while the BSS specific frequency decreases in both clusters
as $r$ increases from 0 to $\sim 4r_c$, the decrease is much
larger in M3. In 47~Tuc it is a factor of 5.5, dropping from
2.64 down to 0.48; in M3 the drop is a factor of 15 (from
2.76 to 0.2). (3) the specific frequency minimum in
47~Tuc appears to be much broader than that observed in M3.
In 47 Tuc the depletion zone extends from $\sim 4 r_c$
to 20--$22~r_c$ with the upturn of the BSS density
occurring at $\sim 25~r_c$, while in M3 the BSS specific
frequency is already rising at $\sim 8 r_c$.
  
F97 argued that the
bimodal distribution of BSS in M3 was  a signature that two formation
scenarios were active in the same cluster, the {\it
external}
BSS  arising from
mass transfer in primordial binaries and the {\it central} 
BSS arising from stellar
interactions which lead to mergers. As earlier noted by
Bailyn \& Pinsonneault (1995), the 
luminosity functions of the two BSS samples 
differ as  theoretically expected for
the two different mechanisms.

\begin{figure}[!ht]
\plotone{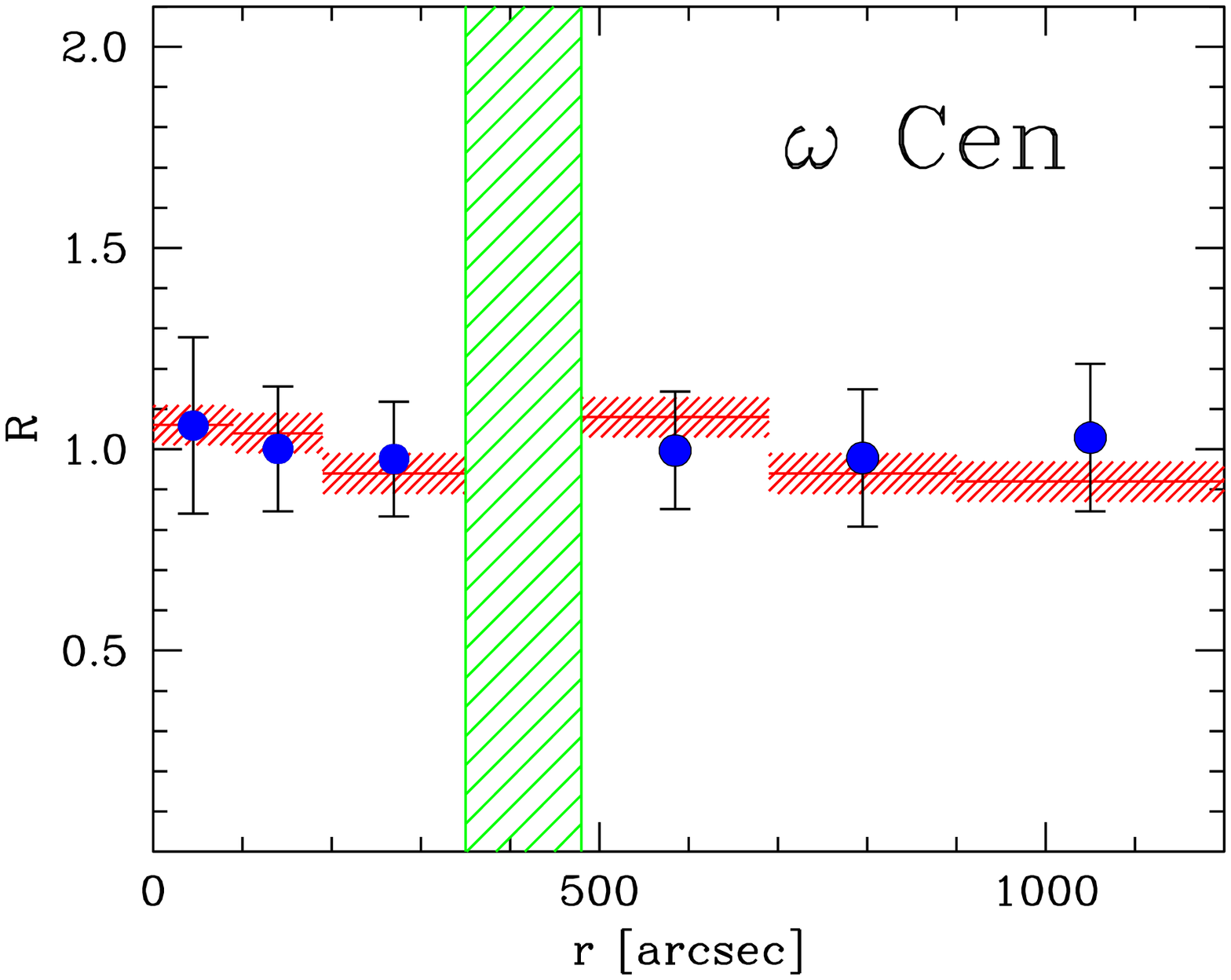}
\caption{The double-normalized relative 
frequency $R_{BSS}$  of the BSS ({\it filled
circles}) in $\omega$ Cen.   The shaded area marks the cluster region 
we excluded in order to avoid incompleteness
problems. The horizontal  lines show  the relative frequency
of the RGB stars used as reference population. (From Ferraro et al 2005).}
\end{figure}

Sigurdsson et al. (1994) offered another explanation
for the bimodal BSS distribution in M3. They suggested that the 
{\it external} BSS
were formed in the core and then ejected into the outer regions by the
recoil from the interactions. Those binaries which get kicked out to a
few core radii ($r_c$) rapidly drift back to the center of the cluster
due to mass segregation, leading to a concentration of BSS near the
center and a paucity of BSS in the outer parts of this region.  More
energetic kicks will take the BSS to larger distances; these stars
require much more time to drift back toward the core and may account
for the overabundance of BSS at large distances.  In order to discern
between different possibilities accurate   simulations
with suitable dynamical codes  
are necessary.  Mapelli et al (2004)
modeled the evolution of BSS in 47 Tuc, mimicking their
 dynamics in a multimass King model, by using  a new version
 of the dynamical code described by Sigurdsson \& Phinney
 (1995). Their results demonstrate that the observed
 spatial distribution cannot be explained within a purely
 collisional scenario in which BSS are generated exclusively
 in the core through stellar interactions. In fact, an accurate
 reproduction of the BSS radial distribution  can be
 obtained only requiring that a sizable fraction of BSS is
 generated in the peripheral regions of the cluster inside
 primordial binaries that evolve in isolation and experiencing
 mass transfer.  
 
A BSS specific upturn similar to that observed in 47 Tuc and
M3 has also been detected in M55 by Zaggia et al.  (1997).
 This result is based on ground-based observations sampling
only a quadrant of the cluster. Since ground-based observations tend
to hide BSS in the central region of the cluster, the bimodality in
M55 could be even stronger than that found by Zaggia et al
(1997). This is of particular significance because M55 has a central
density significantly lower than M3 and 47 Tuc  and could
suggest that the bimodal distribution is not related to the central
density of the parent cluster.  
Moreover a BSS bimodal radial distribution has been also
recently detected in NGC6752 by Sabbi et al (2004).
At present, there are at least 4 GCs in which the
 BSS radial distribution seems to be bimodal: M3, 47 Tuc,
 M55 and NGC6752 (see Figure 9).

Though the number of the
surveyed clusters is low, 
these discoveries suggest that the
{\it peculiar} radial distribution first found in M3 is much more {\it
common} that was thought.  {\it Indeed, it could be the ``natural''
BSS radial distribution}.  Clearly, generalizations cannot be made
from a sample of  a few clusters. Hence we need to
characterize the BSS radial
distribution on a much more solid statistical base. 
 First results from simulations  indicate that 
bimodality  is a signature  that both collisions and
primordial binaries play an important role.

However, there is recent evidence that  BSS do not follow
this rule
in at least one cluster. Using combined (HST+ground-based) observations  
Ferraro et al (2005) selected the largest population of BSS
ever observed in a stellar system in the giant cluster
$\omega$~Cen. Conversely to any other GC surveyed up to now,
 they found that
 the BSS frequency in this stellar system
 {\it does not peak in the center and does not vary 
 with the distance from the cluster center}.
 As can be seen from Figure 10, the double normalized BSS
 ratio turns out to be nicely constant over the entire
 cluster extension and it is fully consistent with the
 reference population. This is the very first time that such
 a trend has been found for BSS.
 This  result is surprising since the relaxation time for the
core of the cluster ($\tau \sim 7$ Gyr) is a factor two lower than the
estimated cluster age ($t \sim 13$ Gyr). Clearly this evidence could be
related to the complex history of this peculiar stellar system (see
recent results in Lee et al 1999; Pancino et al 2000, 2002; Ferraro et al
2002, 2004b). {\it However, our observations give the cleanest evidence
that this cluster has not yet reached the energy equipartition even in
the central core} and further support the use of
BSS as probe of the dynamical cluster evolution.

\begin{figure}[!ht]
  \plotone{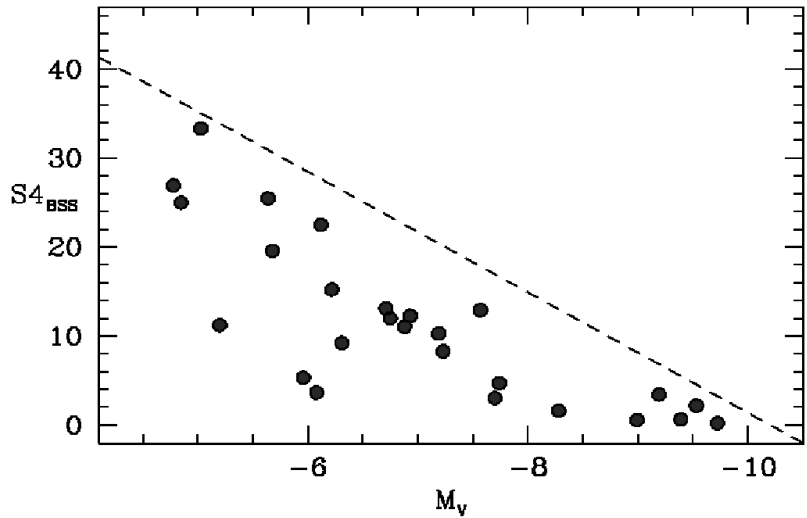}
  \caption{BSS relative frequency ($S4_{BSS}$) as a function of the
  cluster integrated absolute magnitude. (From
  FFB95). }
 \end{figure}
\begin{figure}[!ht]
  \plotone{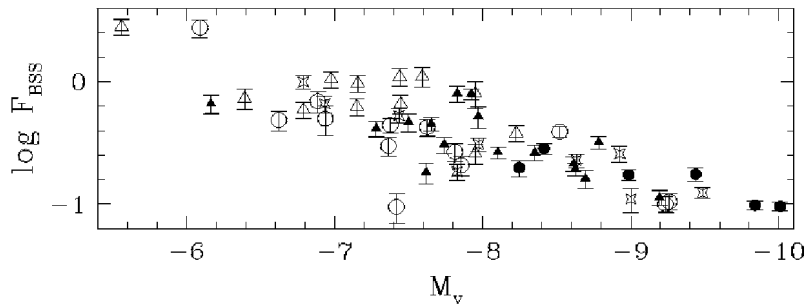}
  \caption{BSS relative frequency ($F_{BSS}$) as a function of the
  cluster integrated absolute magnitude: different symbols
  are used for clusters with different central densities.
  (From Piotto et al 2004).}
 \end{figure}

\subsection{BSS and the parent GC properties}
 
\subsubsection{Specific BSS frequencies and structural parameters}
 
The collection of an homogeneous dataset of BSS in different
clusters allows  a direct
cluster-to-cluster comparison. However, it should be emphasized that
the interpretation of BSS specific frequency in terms of structural
cluster parameters can be risky (see  FFB95). 
In particular, one should keep in mind 
the  intrinsic dependence of any BSS specific frequency on the cluster
luminosity.
 
Recently, Piotto et al. (2004) noted a correlation between BSS specific
frequency and cluster absolute magnitude. The same correlation had
been discussed by FFB95 (see Fig.~11 and 12), 
who showed that much of this effect arose from the 
normalization of the population.
Indeed such a trend  could be generated by the
correlation between the sampled  and the total
luminosity as shown in Fig.~4 of FFB95.  A similar analysis should
be performed on the Piotto et al (2004) sample in order to establish
the role of the normalization factor in defining the observed
trend.  

Based on the  results of Piotto et al., Davies et al. (2004)
developed a model for the production of BSS in GGCs. In the low mass
systems ($M_V > -8$) BSS arise mostly from mass exchange in primordial
binaries. In more massive systems collisions produce mergers of the
primordial binaries early in the cluster history. BSS resulting from
these mergers long ago evolved away. Once the primordial binaries were
used up, BSS produced via this channel disappeared. In the cores of the
most massive systems ($M_V < -9$) collisional BSS are
produced (see Fig.~6 of Davies et al.). The  
 working hypothesis proposed by Davies et al (2004) is 
 interesting, however
detailed cluster-to-cluster comparison has shown that the scenario is
much more complex than that, since the
dynamical history of each cluster apparently plays a 
significant role in
determining the origin and radial distribution BSS content
(see next Section).

\subsubsection{Cluster to cluster comparison}

In the contest of a direct comparison of the BSS content 
in different clusters, particularly interesting is the case
of M80, which shows an exceptionally high BSS content:
more  than 300 BSS have been discovered  (Ferraro et al
1999).  This is among the largest and most  concentrated
BSS population ever found in a GGC (see Figure 13). Indeed
only the largest stellar system in the Halo, $\omega$
Centauri, has been found to harbor a   BSS population
(Ferraro et al 2005) larger than  that discovered in M80. 
M80 is the  GGC which has the largest central density among
those  not core-collapsed yet. However, the stellar density
cannot explain  such a large population, since other
clusters with similar central density  harbor much fewer
BSS (see the case of 47 Tuc, Ferraro et al 2001a,  NGC6388
Piotto et al 2003). Ferraro et al (1999) suggested that M80
is in a transient dynamical state during which stellar
interactions  are delaying the core-collapse process,
leading to an exceptionally  large population of
collisional BSS. If this hypothesis would be further
confirmed,  this discovery could be the first direct
evidence  that stellar collisions could indeed be effective
in delaying the core collapse.

Interesting enough, clusters that have already experienced
(or are experiencing) the collapse of the core show a small
BSS population. Indeed this is the case of NGC6752. This
cluster has been recently found (see Ferraro et al 2003d)
to be dynamically evolved  probably undergoing a Post Core
Collapse bounce (see Section 3.4). A recent search for BSS
in the central region of this cluster indicated a
surprisingly low   BSS content: the specific number of BSS
is among the lowest ever  measured in a  GC (Sabbi et al
2004, see Figure 14).

\begin{figure}[!ht]
\plotone{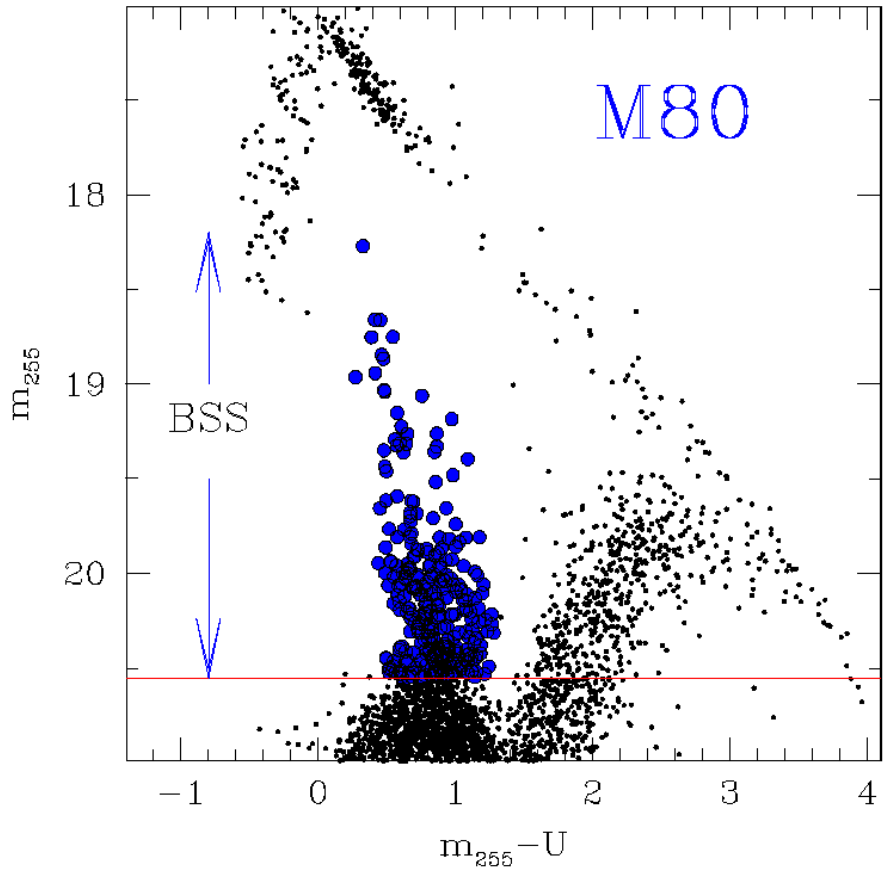}
\caption{ (m$_{255}$, m$_{255}$-m$_{336}$) CMD for the central region
of M80, from WFPC2/HST observations.  The 
300 BSS candidates  discovered in
this cluster are marked as filled
circles.
(From Ferraro et al 1999).}
\end{figure}

\begin{figure}[!ht]
\plotone{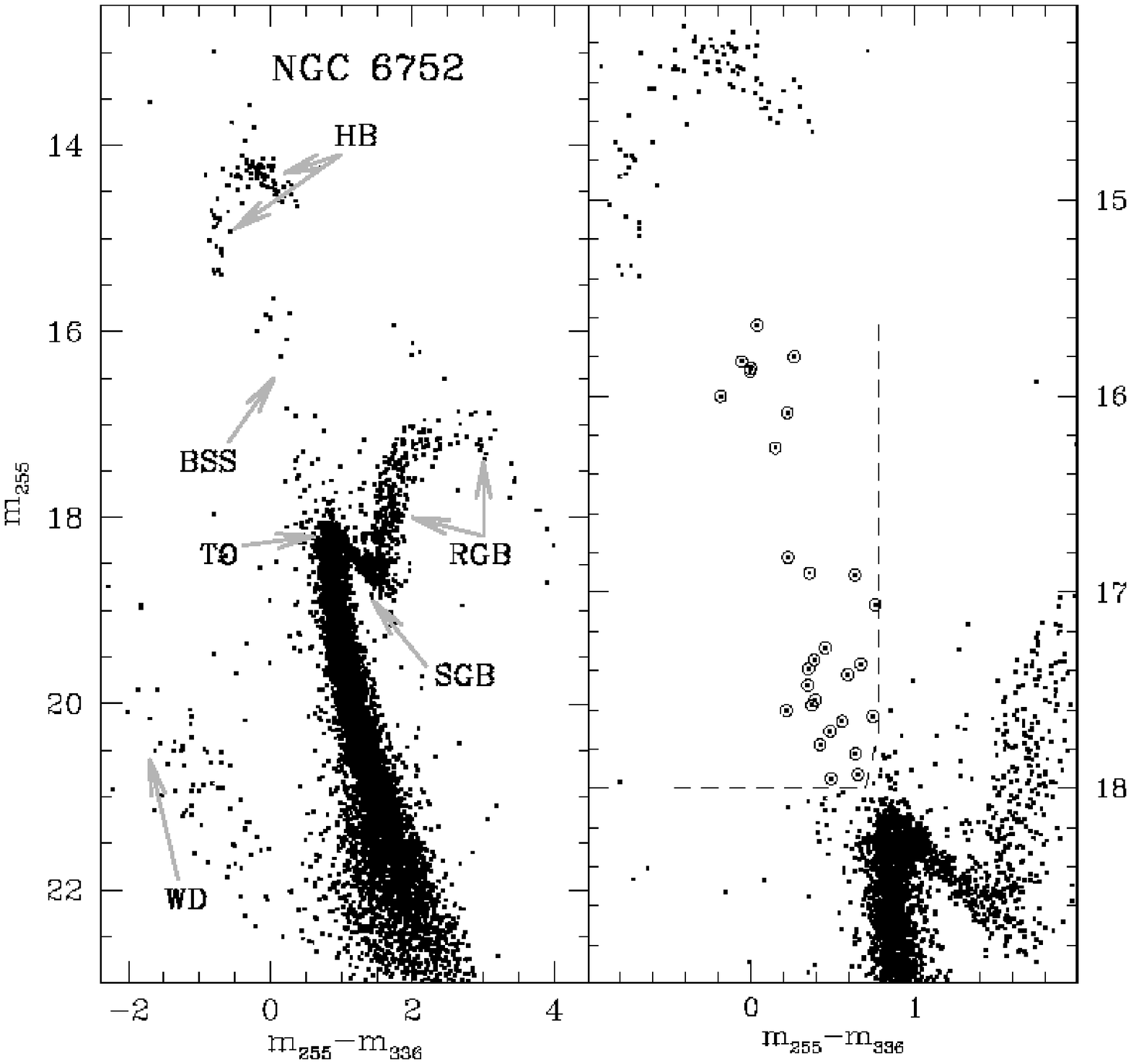}
\caption{ (m$_{255}$, m$_{255}$-m$_{336}$) CMD for the central region
of NGC~6752, from WFPC2/HST observations. {\it Left panel:} The whole CMD. The
main branches are indicated. {\it Right panel:} The zoomed CMD in the BSS
region. The selected BSS are marked with large empty circles.
(From Sabbi et al 2004).}
\end{figure}

F03 presented a direct comparison  of the BSS content for 6
clusters observed in the UV with HST (namely M3, M13, M80,
NGC288, M92 and M10). Figure 15 shows the
($m_{255},~m_{255}-m_{336}$) CMDs for these   clusters.
More than 50,000 stars are plotted in the six panels of
Figure 15. The CMD of each cluster has been shifted to 
match that of M3  using the brightest portion of the HB as
the normalization region. The solid horizontal line (at
$m_{255}=19$) in the figure marks the threshold magnitude
for the selection of the bright (hereafter bBSS) sample.
Such a dataset allows a direct cluster-to-cluster
comparison. A number of interesting results have been
obtained and in the following we just
briefly discuss the two major ones:

\begin{itemize}

\item  the
specific frequency of BSS largely varies from cluster to
cluster. The specific frequency of BSS compared to the
number of  HB stars varies from 0.07 to 0.92 for these six 
clusters, and does not seem to be correlated with central
density,  total mass, velocity dispersion, or any other
obvious cluster  property. Twins clusters as M3 and M13
harbor a quite different BSS population: the specific
frequency in M13 is the lowest ever measured in a GC
(0.07), and it  turns out to be 4 times lower than that
measured in M3 (0.28). Which is the origin of this
difference?  The paucity of BSS in M13 suggests that 
either the primordial population of binaries in M13 was
poor or that  most of them were destroyed.  Alternatively,
as suggested by F97, the mechanism producing BSS in the
central region of M3 is  more efficient than M13 because M3
and M13 are in different dynamical  evolutionary phases.  

\begin{figure}[!ht]
\plotone{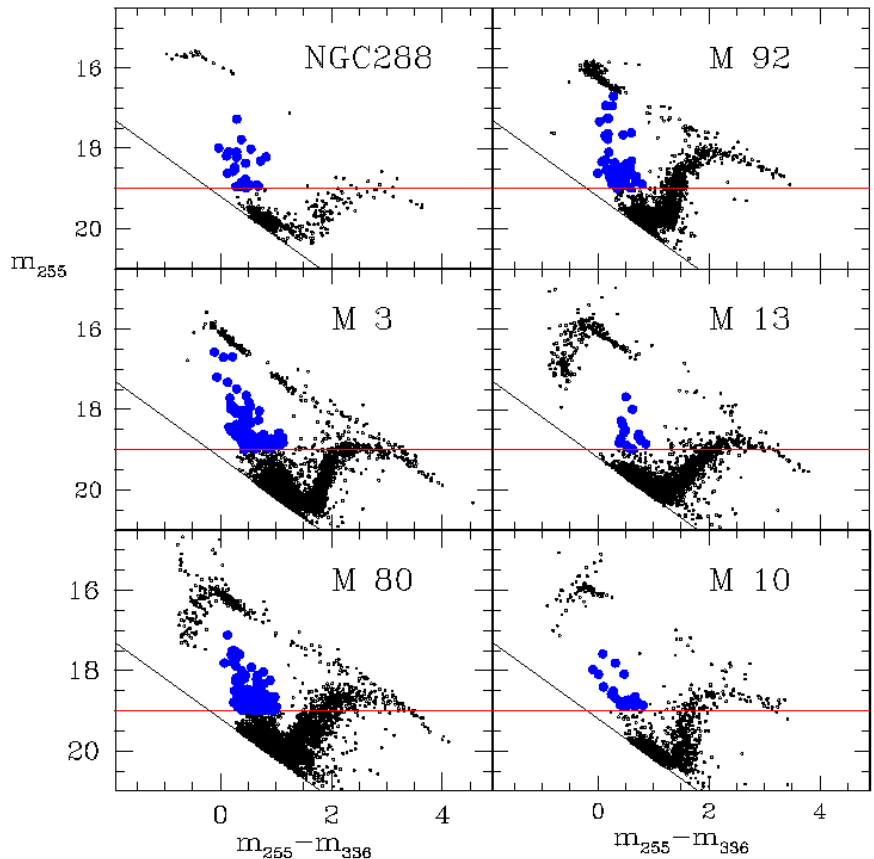}
\caption{($m_{255}, m_{255}-m_{336}$) CMDs for the selected clusters. 
Horizontal and vertical shifts have been applied to all   CMDs in order  
to match the main sequences of M3.  
The horizontal solid line corresponds to $m_{255}=19$ in M3. 
The bright BSS candidates are marked as large filled
circles. (From F03).   
}
\end{figure}

\begin{figure}[!ht]
\plotone{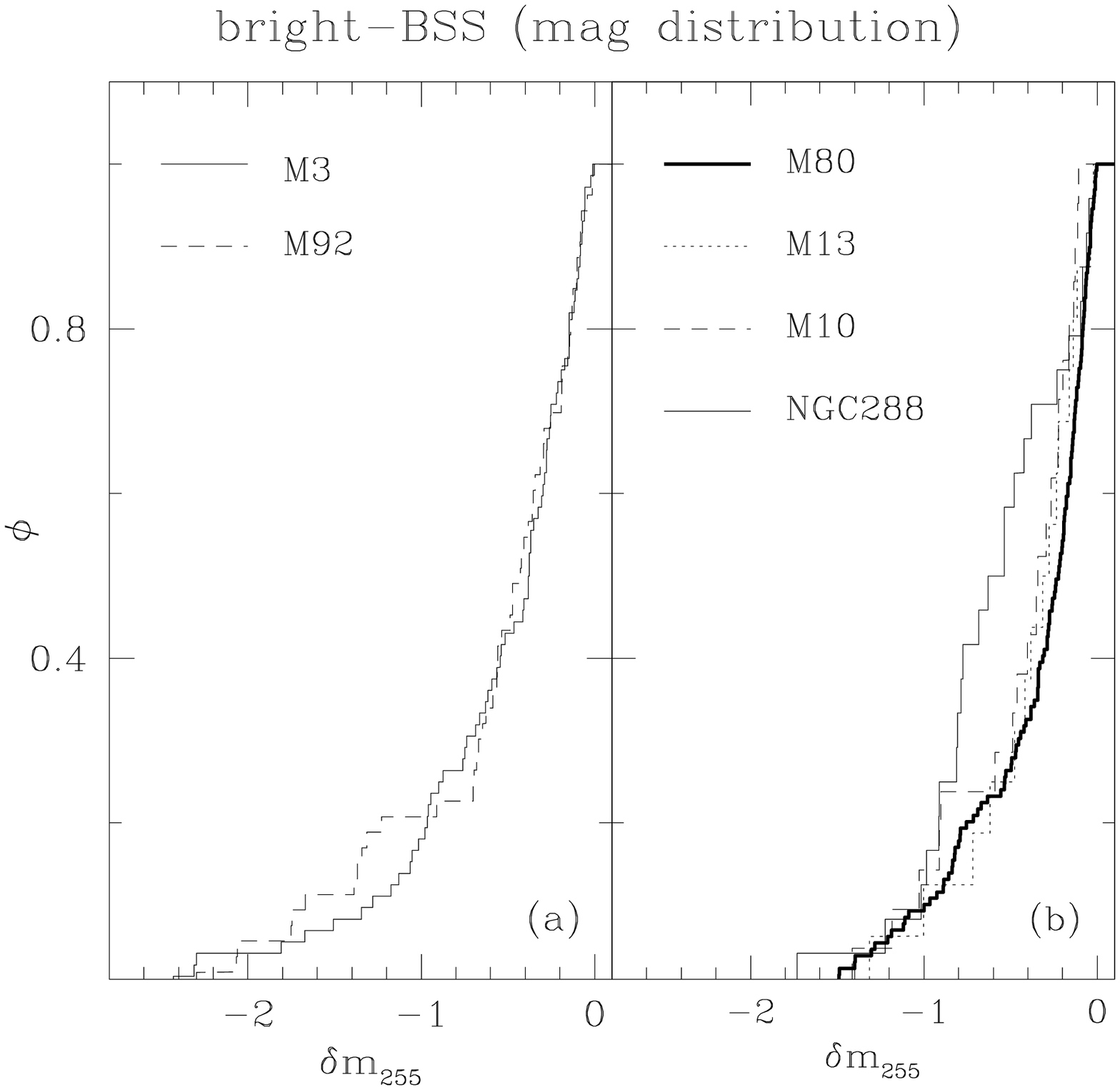} 
\caption{Cumulative  magnitude distributions for the {\it bright} 
 BSS   for  the 6 clusters presented by F03. 
 The $\delta m_{255}$ parameter is the 
 difference in magnitude with respect to the limit threshold 
  ($m_{255}=19$) for bright-BSS. 
 In {\it Panel (a)} the BSS distributions 
 for  M3 and M92 (the two clusters for which the  
 BSS distribution extends up to more than  
 two magnitudes brighter than the threshold) are compared. 
 In  {\it Panel (b)} the BSS magnitude distributions for the other 4 
  clusters are plotted.   (From F03).   
}
\end{figure}

In this respect, the most surprising result is that the two
clusters with the largest BSS specific frequency are at the
central density extremes of our sample: NGC 288 (lowest
central density) and M80 (highest). The BSS specific
frequency measured in these clusters (Bellazzini et al
2002, and Ferraro et al 1999) suggest that  both of them
have almost as many BSS as HB stars in the central region.
F03 have shown that   the  collision channel is more than
one order of magnitude more efficient in a dense cluster
like M80 than in a low density cluster like NGC 288.
Therefore, the high specific frequency of BSS in NGC 288
suggests that the binary fraction in this cluster is much
higher than the one in M80. Only then would one
expect a similar encounter frequency in the two clusters.
Note that a cluster like M80 may have originally had a
higher binary fraction but because of the efficiency of
encounters, those primordial binaries were ``used up''
early in the history of the cluster, producing some
collisional BSS which  have evolved away from the MS (see
for example the evolved BSS population found in M3 and
M80 by F97 and Ferraro et al. (1999, see also Section 2.6).

However, without invoking {\it ad hoc} binary content, 
{\it a more natural explanation} for the origin of BSS in
NGC 288 (as discussed in Bellazzini et al (2002)) is the
mass transfer process in primordial binary systems (Carney
et al 2001). We also know that the binary fraction in the
core is   $\sim 10$ to 38\% (Bellazzini et al 2002), so BSS
formed in this low density cluster should be the result of
binary evolution rather than stellar collisions.  Here we
probably have another confirmation of the scenario
suggested by Fusi Pecci et al (1992, and references
therein): BSS living in different environments have
different origins. In this case the above result
demonstrated that both channels are quite efficient in
producing BSS.

\item  
In Figure 16 the magnitude distributions (equivalent to a
Luminosity Function) 
of bBSS for the six clusters are compared.  In doing
this we use the parameter $\delta m_{255}$ defined as the magnitude of
each bBSS  with respect to the
magnitude threshold (assumed at $m_{255}=19$ - see Figure
15): then
$\delta m_{255}= m^{bBSS}_{255}-19.0$. From the comparison shown in
Figure 16 ({\it panel(a)}) the bBSS magnitude distributions for M3 and
M92 appear to be quite similar and both are significantly different from
those obtained in the other clusters. This is essentially because in
both clusters the bBSS magnitude distribution seems to have
a tail extending to brighter magnitudes (the bBSS magnitude tip
reaches $\delta m_{255}\sim -2.5$). A KS test applied to these two
distributions yields a probability of $93$\% that they are extracted
from the same distribution. In {\it panel(b)} we see that the bBSS
magnitude distribution of M13, M10 and M80 are essentially
indistinguishable from each other and significantly different from M3
and M92. A KS test applied to the three LFs confirms that they
are extracted from the same parent distribution. Moreover, a KS test
applied to the total LFs obtained by combining the data for the two
groups: M3 and M92 ({\it group(a)}), and M13, M80 and M10 ({\it
group(b)}) shows that the the bBSS-LFs of {\it group(a)} and 
{\it group(b)} are
not compatible (at $3\sigma$ level).

It is interesting to note that the clusters grouped on the basis of
bBSS-LFs have some similarities in their HB morphology. The three
clusters of {\it group(b)} have an extended HB blue tail; the two clusters
of {\it group(a)} have no HB extension. Could there be a connection between
the bBSS photometric properties and the HB morphology? This
possibility needs to be further investigated.

\end{itemize}

 \begin{figure}[!ht]
\plotone{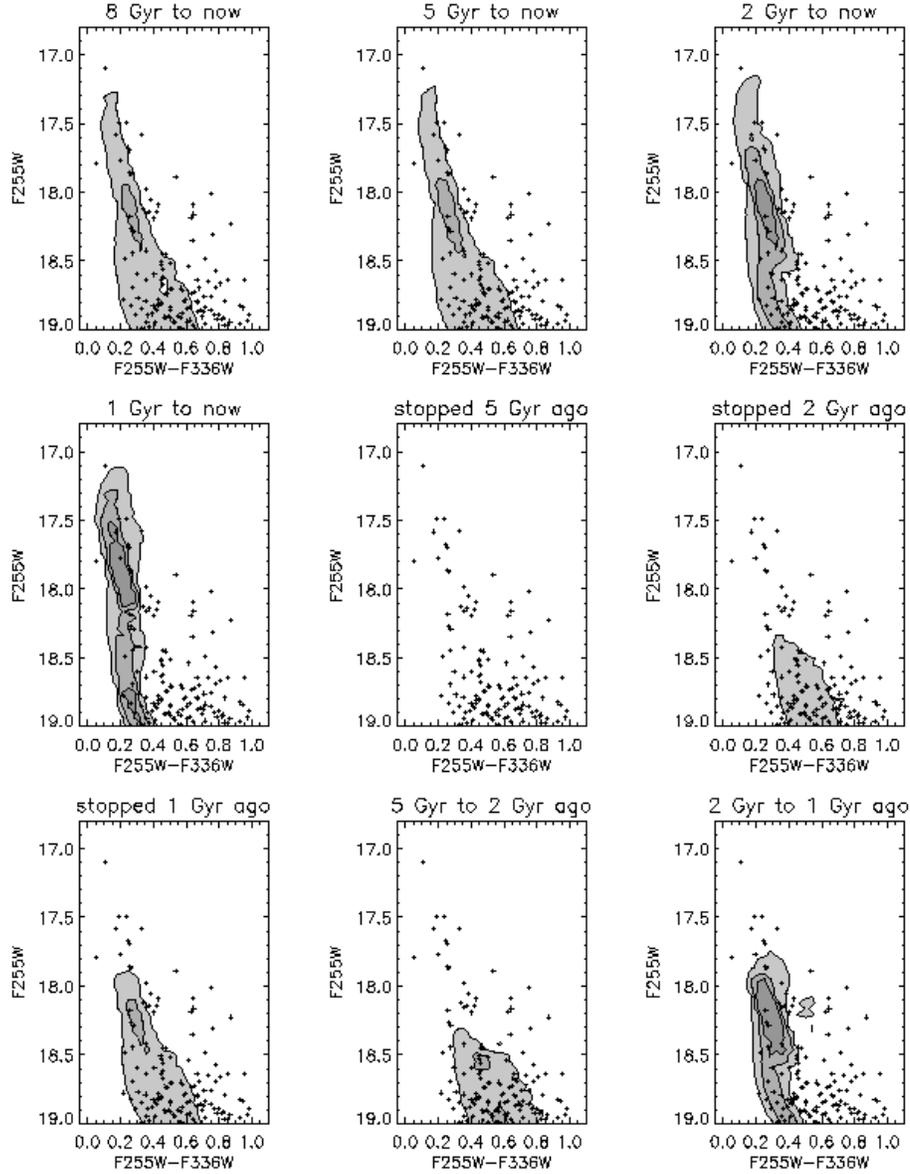}
\caption{The distribution of bright BSS 
in the CMD of M80, compared to theoretical 
distributions. The observations are plotted as crosses, and the 
grey scale contours give the theoretical distributions, with darker 
colors indicating more BSS. The different panels 
correspond to different eras of constant BSS formation, as 
indicated at the top of each panel. (From F03).   
}
\end{figure}

\subsection{Comparison with collisional models} 

A preliminary comparison between the observed BSS distribution 
in the CMD (and their LF) with those obtained from 
collisional models have been presented in F03. 
The  models used there are
described in detail in Sills \& Bailyn (1999). The main assumption
is that all the BSS
in the central regions are  {\it formed through
stellar collisions  during an encounter between a
single star and a binary system}. Unfortunately,  
while binary-binary collisions may
well be important, we do not currently have the capability of
modeling the BSS they might produce.

The other ingredients of the collisional models are discussed
in F03,
in the following we list the major assumptions: 

\begin{itemize}

\item  The trajectories of the stars during the collision are modeled using
the STARLAB software package (McMillan \& Hut 1996).

\item The adopted  mass function has an index $x=-2$, and the
mass distribution within the binary systems are drawn from a Salpeter
mass function ($x=1.35$).

\item The adopted binary fraction is 20\% with  a
binary period distribution which is flat in $\log P$.

\item The total
stellar density is derived from the central density of each
cluster.

\end{itemize}

In order to explore the effects of non-constant BSS formation rates, we 
considered a series of truncated rates, 
namely  constant for some portions of the 
cluster lifetime, and zero otherwise.  This assumption is obviously 
unphysical---the relevant encounter rates would presumably change 
smoothly on timescales comparable to the relaxation time.  However 
these models do demonstrate how the distribution of BSS in the 
CMD depends on when the BSS were created, and 
thus provide a basis for understanding more complicated and realistic 
formation rates.

Figure 17 shows an example of the  distribution of
BSS in the $m_{255},~m_{255}-m_{336}$ CMD
as predicted by collisional models in the case of M80.  
 In each  panel, a
different assumption for the BSS formation rate has been
adopted. The differences between the models can be understood
in terms of lifetimes of the individual collision products which make
up the distributions. For example, if BSS production stopped 5 Gyr
ago (central panel), we predict that there should be no observed bBSS at
present because all the massive BSS  had time to evolve off
to the RGB. At the other extreme, if all the BSS were formed
in the last Gyr (panel marked ``1 Gyr to now''), only a few
(if any) of the 
brightest (i.e. most massive) BSS  started moving towards
the subgiant branch. 

As a general result, F03
concluded that the 
BSS formation occurred in all clusters over 
last 5 Gyr at least, and more were formed 2--5 Gyr ago than in the 
recent past. Of course this result needs to be tested using models of GC 
dynamical evolution in which the feedback between stellar collisions and 
cluster evolution is modeled explicitly. Our assumption of a BSS 
formation rate which is either constant or zero is unphysical, and 
more complicated models are clearly required.

\subsection{Evolved BSS on the HB}

Renzini \& Fusi Pecci
(1988) suggested to search for Evolved BSS (hereafter E-BSS)
 during their core helium burning
phase since they should appear to be redder and brighter than {\it
normal} HB stars.  Following this prescription, Fusi Pecci et
al. (1992) identified a few E-BSS candidates in several clusters with
predominantly blue HBs, where the likelihood of confusing E-BSS stars
with true HB or evolved HB stars was minimized. Because of the small
numbers there is always the possibility that some or even most of these
candidate E-BSS are due to field contamination. However, 
near the cluster centers,
field contamination should be less of a problem. Following
Fusi Pecci et al (1992), F97
identified  a sample of 19 E-BSS candidates in M3  (see Figure
18) and argued that the radial distribution of E-BSS was similar to that
of the BSS. 
 
\begin{figure}[!ht]
  \plotone{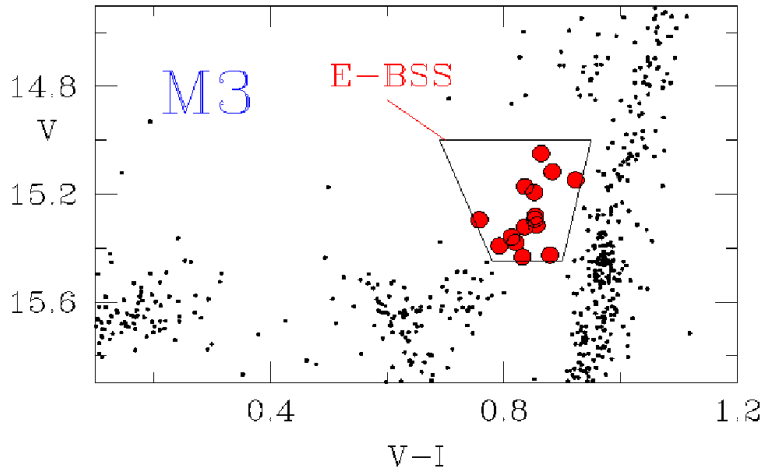}
  \caption{ 
  Zoomed $(V,~V-I)-$CMD
of the HB region of M3. Evolved BSS candidates 
are plotted as large filled circles. 
(From F97).}
 \end{figure}

\begin{figure}[!ht]
  \plotone{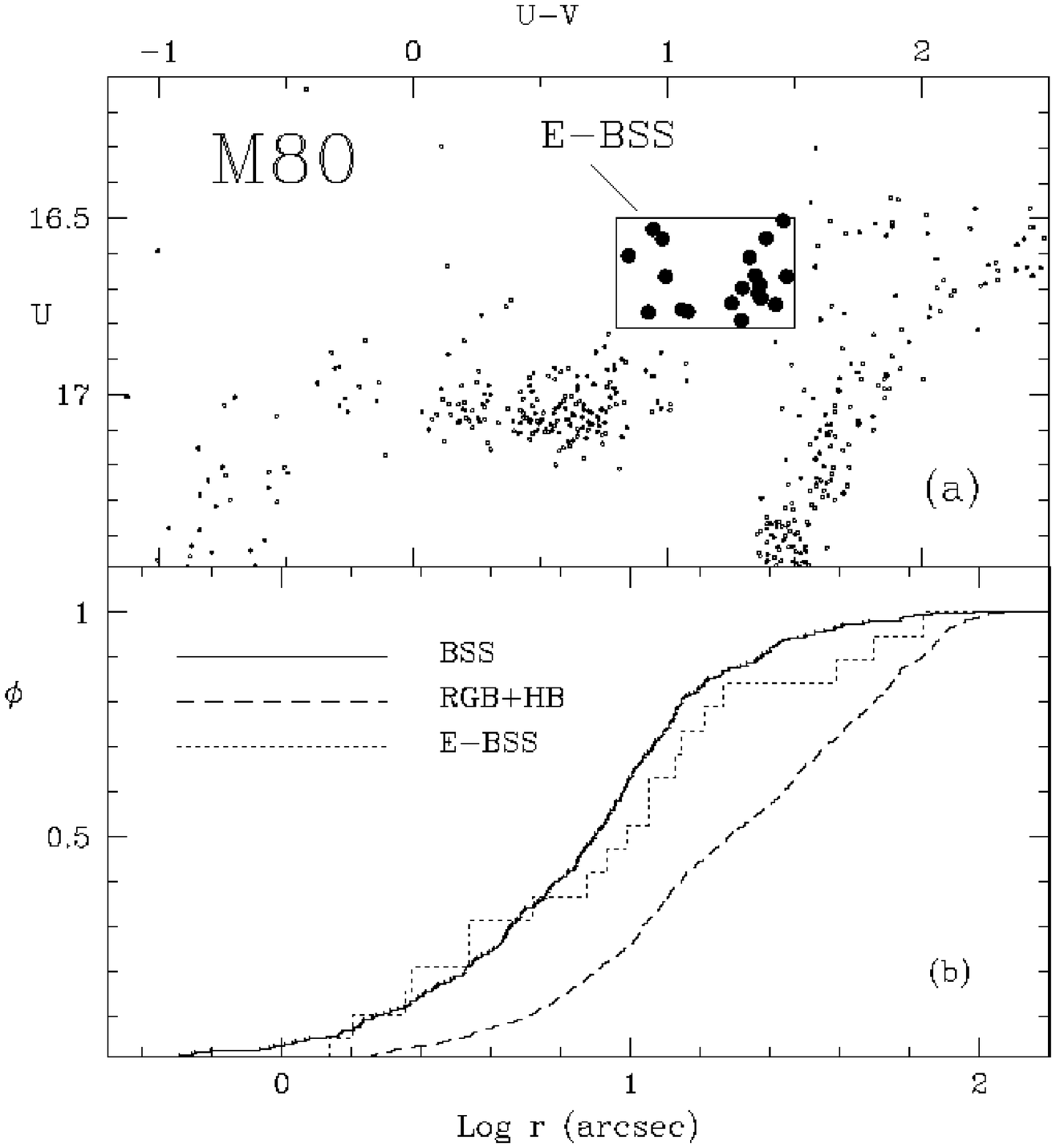}
  \caption{  {\it Panel (a)} 
  Zoomed $(U,~U-V)$CMD
of the HB region of M80. Evolved BSS (E-BSS) candidates
are plotted as large filled circles. 
{\it Panel (b)} Cumulative radial distribution ($\phi$) for BSS
(heavy solid line), E-BSS (dotted line)
compared to the HB+RGB stars (dashed line) as a function
of their projected distance ($r$) from the cluster center. 
(From Ferraro et al 1999).}
 \end{figure}

The large population of BSS discovered by Ferraro et al
(1999) in
M80 allowed a further possibility to search for E-BSS, in
fact, M80 offers some  advantages over M3 in selecting  E-BSS:
(1) it has a very blue HB, so there should be less confusion between red
HB stars and E-BSS; (2) it has a larger number of BSS.
  In Figure 19 {\it Panel (a)}  a zoomed $(U,~U-V)$ CMD of the HB
region is shown and the expected location for E-BSS 
has been indicated as a box;
19 E-BSS (plotted as large filled circles) lie in the box.  
The cumulative radial distribution of the E-BSS stars
shown in {\it Panel (b)} of Figure 19 is 
   quite similar to the BSS
distribution and significantly different from that of the HB-RGB.  A
Kolmogorov-Smirnov test shows that the probability that the E-BSS and
BSS populations have been extracted from the same distribution
is $\sim 67$\%, while the probability that the E-BSS and the
RGB-HB population have the  same distribution is
only $\sim 1.6 $\%. This result confirms the expectation that the 
E-BSS share the same distribution of the BSS and they are
both a more massive population than the bulk of the
cluster stars.  

Under the assumption that the connection between the E-BSS and
the BSS population is real, one can relate the population ratios
and the lifetimes of these evolutionary stages. 
Earlier studies (Fusi Pecci et al 1992) have suggested that the ratio of
bright BSS (b-BSS) to E-BSS is 
$N_{\rm b-BSS}/N_{\rm E-BSS} \approx 6.5$. Interesting
enough, the population ratios are quite similar in both
clusters: in M3 F97 counted 19 E-BSS over a population of
122 bBSS; the same number of E-BSS candidates 
 have been
selected by Ferraro et al (1999) in M80 compared with
a population of 129 b-BSS. The population ratio 
turns out to be $N_{\rm b-BSS}/N_{\rm
E-BSS} = 6.6$, fully consistent with earlier
studies. 
Moreover because both  BSS and E-BSS samples are so cleanly defined
in the case of M80,
the ratio of the total number of BSS to E-BSS, $N_{\rm BSS} /N_{\rm
E-BSS} \sim 16$, should be useful in testing lifetimes of
BSS models.

\section{MSP companions in GCs}

Among the possible collisional by-product zoo,  
MSPs are invaluable probes to study cluster dynamics: they 
form in binary systems containing a neutron star (NS) which is
eventually spun up through mass accretion from the evolving 
companion.
Despite the large difference in total mass between the disk 
of the Galaxy and the GC system, about $50$\% of the entire MSP 
population has been found in the latter.  
This is not surprising since in the Galactic disk 
MSPs can only form through the evolution of primordial binaries, 
while in GC cores
dynamical interactions can lead to the formation 
of several different  binary systems, suitable for recycling 
NS.

The search for optical counterpart to MSP companion in GCs
is a  relatively recent branch of this research, 
since the first identification has been done only a few
years ago in the core of 47 Tuc. Edmonds et al (2001)
identified $U_{opt}$,
 the companion to PSR~J0024$-$7203U: this object 
 turned out to be  a faint blue variable
whose position in the CMD  is consistent a cooling
helium WD. This is fully in agreement with the MSP recycling
scenario, where the usual companion to a binary
MSP is an exhausted star.

\subsection{The surprising companion to the  MSP
 in NGC6397 }

A major surprise came from the
 optical identification  of
the companion to the binary MSP
PSR J1740-5340 in   NGC6397.

PSR J1740-5340 was identified by D'Amico et al. (2001a) 
during a systematic search for MSP in GCs  carried out 
with the Parkes radio telescope.  
The pulsar displays eclipses at a frequency of 1.4 GHz for more 
than 40\% of the 32.5 hr orbital period and exhibits striking
irregularities of the radio signal at all orbital phases
(D'Amico et al 2001b). 
This suggests that the MSP is orbiting within a large envelope of
matter released from the companion, whose
interaction with the pulsar wind could be responsible for the
modulated and probably extended X-ray emission detected with  
CHANDRA (Grindlay et al. 2001, 2002).  By using 
high resolution multiband HST observations  and 
the position of the
MSP inferred from radio timing,  Ferraro et al (2001b) identified  
a bright variable star (hereafter COM J1740-5340)
as the optical counterpart to the MSP companion,
whose optical modulation nicely agrees 
with the orbital period of the MSP itself. 
The optical counterpart shows a quite anomalous position in
the CMD since it is located at the luminosity of the TO 
point  but it has an anomalous red color (see Figure
20).

\begin{figure}[!ht]
  \plotone{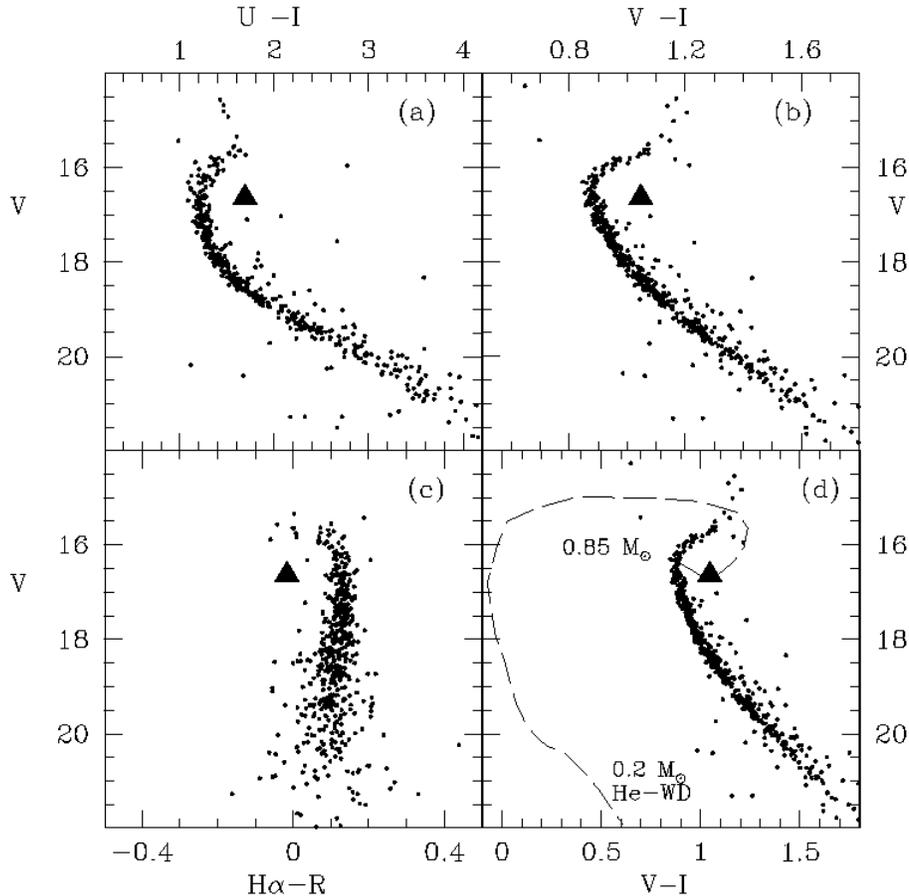}
  \caption{ The optical counterpart to the MSP companion in
  NGC6397. Multiband Color Magnitude Diagram for stars
  detected in a region  around the nominal position of the
   MSP. The optical counterpart of the MSP companion is
   marked with a large filled triangle in all panels. 
   In {\it Panel (d)} the
    evolution path suggested by Burderi et al.
    (2002) is shown. (From Ferraro et al (2001b).}
 \end{figure}
 
As quoted above, in the classical framework of the MSP recycling scenario,
the usual companion to a binary MSP should be either a WD
or a very light ($0.01-0.03~M_{\odot}$), almost exhausted star. 
None of these scenarios can be applied to COM J1740-5340:
it is too luminous to be a WD ($V\sim16.5$, comparable to
the TO stars of NGC6397); moreover its mass ($M\sim
0.2 M_{\odot}$), recently
constrained by radial velocity observations (see Figure 21),
is too high for a very light stellar
companion.
As a consequence, a wealth of intriguing
scenarios have flourished in order to explain the nature of this binary (see
Orosz \& van Kerkwijk 2003, Grindlay et al. 2002 for a review).
In particular, Burderi et al. (2002) suggested that the
position of  COM J1740-5340 in the CMD
(CMD) is consistent with
the evolution of an (slightly) evolved
Sub Giant Branch (SGB) star orbiting the NS and  loosing mass. The
 future evolution of this system will generate a He-WD/MSP
 pair (see {\it panel (d)} in Figure 20).
 COM J1740-5340 could be a   star acquired 
by exchange interaction in the cluster core or alternatively 
the same star that spun up the MSP and still 
overflowing its Roche lobe.
The latter case suggests the fascinating possibility  
that PSR J1740-5340 is a new-born MSP, the very first  
observed just after the end of the recycling process.

This is the first example ever observed of a MSP companion whose 
light curve  is dominated by ellipsoidal variations, suggestive 
of a tidally distorted star, which almost completely fills 
(and is still overflowing) its Roche lobe. 
Thanks to the unusual brightness of the companion ($V\sim
16.5$),  this system represents an unique laboratory to study the
formation mechanism of   binary MSP   in GCs,
allowing unprecedented detailed
spectroscopic observations. In this contest, our group is coordinating 
a spectro-photometric programme at ESO-Telescopes.
In particular a first set of high resolution spectra have
been acquired at the {\it Very Large Telescope} (VLT) with 
UVES. From these  data we
obtained a number of interesting results:

\begin{figure}[!ht]
  \plotone{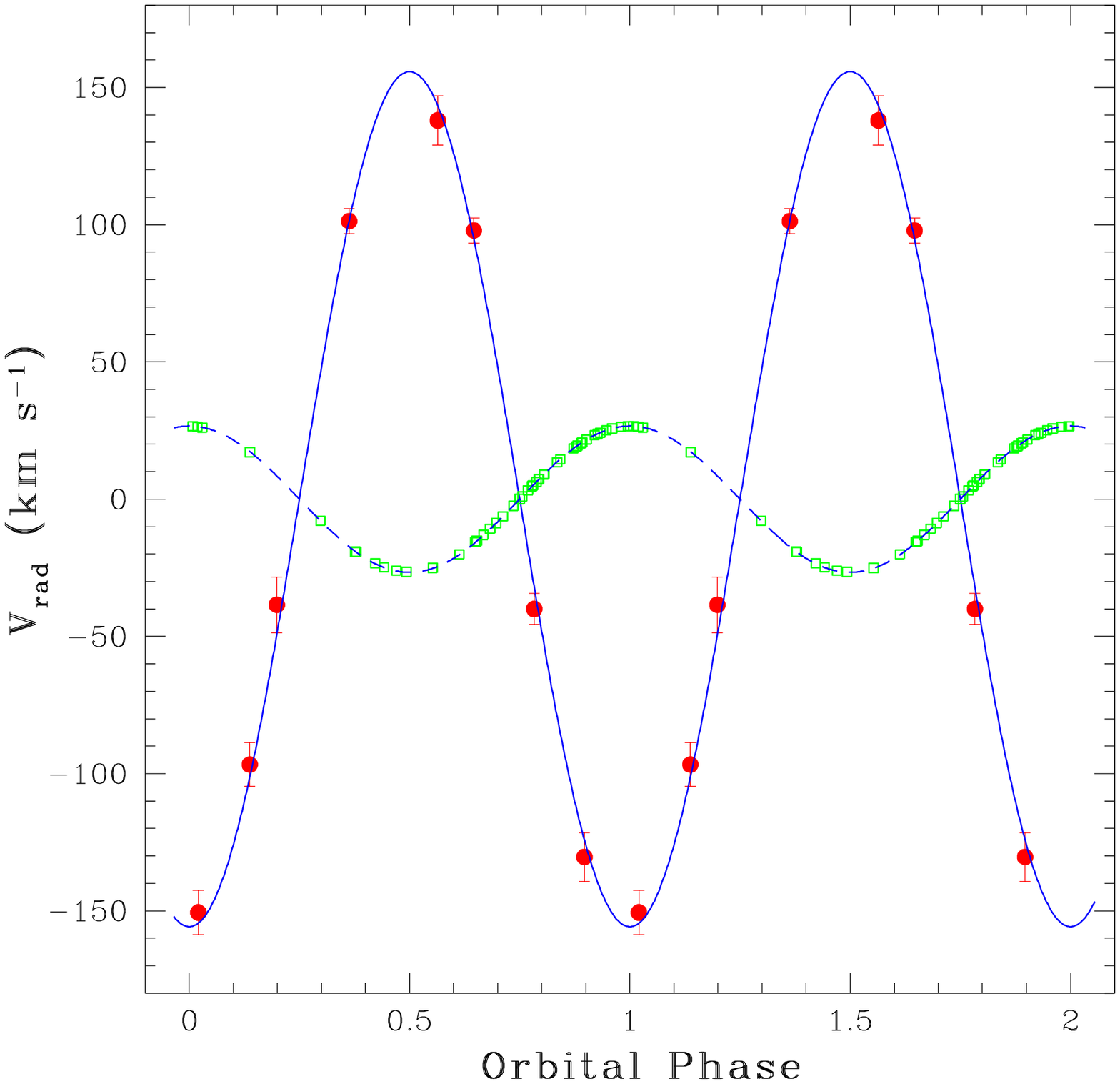}
  \caption{The {\it large dots} are the radial velocity determinations
   for COM J1740-5340. The
{\it solid} line represents the best-fit sinusoidal curve. 
The {\it small open squares} are the radial velocity determinations for
PSR J1740-5340  derived from timing measurements 
and the radio ephemeris (D'Amico et
al. 2001b).  The {\it dotted} line represents the fitted velocity
curve of the pulsar. (From Ferraro et al 2003b).}
 \end{figure}

\begin{figure}[!ht]
  \plotone{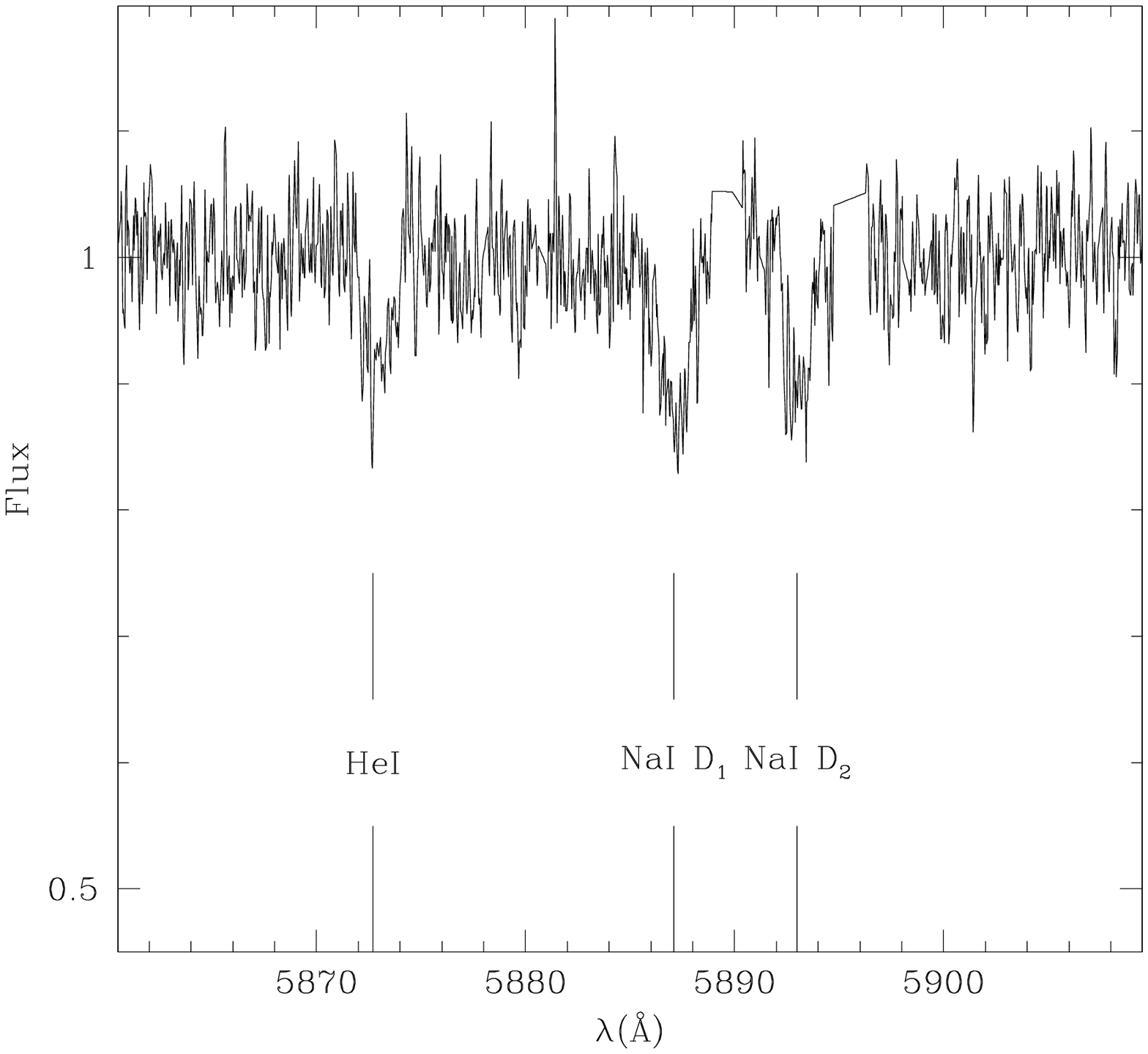}
  \caption{ Portion of the normalized spectrum 
  taken at $\phi=0.02$ in the Na{\sc
i} D lines region (the telluric Na{\sc i} lines have been removed for
clarity).  The spectrum has been smoothed with a boxcar of 3 pixels,
and not shifted to rest wavelengths; the observed RV is $-145.4$ km
s$^{-1}$.  The He{\sc i} absorption line at 5875.6 \AA ~is clearly
visible; Na{\sc i} D lines and the strong He line are also 
indicated.  (From Ferraro et al 2003b). }
 \end{figure}

\begin{itemize}
\item The determination of the radial velocity curve (see
Figure 21) 
allowed an accurate measure of the mass ratio of the system
($M_{PRS}/M_{COM}=5.83\pm0.13$) which suggests a mass of 
$M_{COM}\sim 0.25 M_{\odot}$ by assuming
$M_{PSR}\sim 1.4 M_{\odot}$. (Ferraro et al 2003b).

\item The $H_{\alpha}$ emission from the system was already
noted by Ferraro et al (2001b, see also {\it Panel (c)} in
Figure 20) and fully confirmed by the high-resolution spectra
(Sabbi et al 2003a).
 In particular, the complex structure of the $H_{\alpha}$ line
   suggests the presence of a matter  stream 
 exaping from the companion towards the NS. Note that 
 because of the radiation flux from the pulsar, the
 material would never reach the NS surface, creating 
 a cometary-like gaseous tail which  feeds the presence
 of  (optically thin)  hydrogen gas
 outside the Roche lobe.
 
 \item  The unexpected detection of  strong He I
 absorption lines (see Figure 22) implies the existence of a region
 at $T>10,000K$, significantly hotter than the rest of the
 star (Ferraro et al 2003b, Sabbi et al 2003a,b).
 The
intensity of the He I line correlates with the orbital phase,
suggesting the presence of a region on the companion
surface, heated by the millisecond pulsar flux.

\item  COM J1740-5340  has been found to show a large
rotation velocity ($V sin i = 50\pm 1 Km s^{-1}$).
The derived abundances are found
fully consistent with those of normal unperturbed stars in
NGC 6397, with the exception of a few elements (Li, Ca, and
C). In particular, the lack of C suggests that the star has been peeled
down to regions where incomplete CNO burning occurs (Sabbi
et al. 2003b),
favoring a scenario where the companion is a  SGB star
which has lost most of its mass (see also Ergma \& Sarna
2003).

\end{itemize}

\subsection{The case of NGC6752}

Another  interesting object has captured  our
attention: the binary MSP 
PSR J1911-5958A (hereafter PSR-A), 
recently discovered in the
outskirts of the nearby GC NGC 6752 (D'Amico
et al. 2001a). It is located quite far away (at
about $6'$) from the cluster optical center. 
Indeed PSR-A is the more off-centered pulsar among the sample of 
41 MSPs with known positions in the parent cluster.

\begin{figure}[ht!]
\plotone{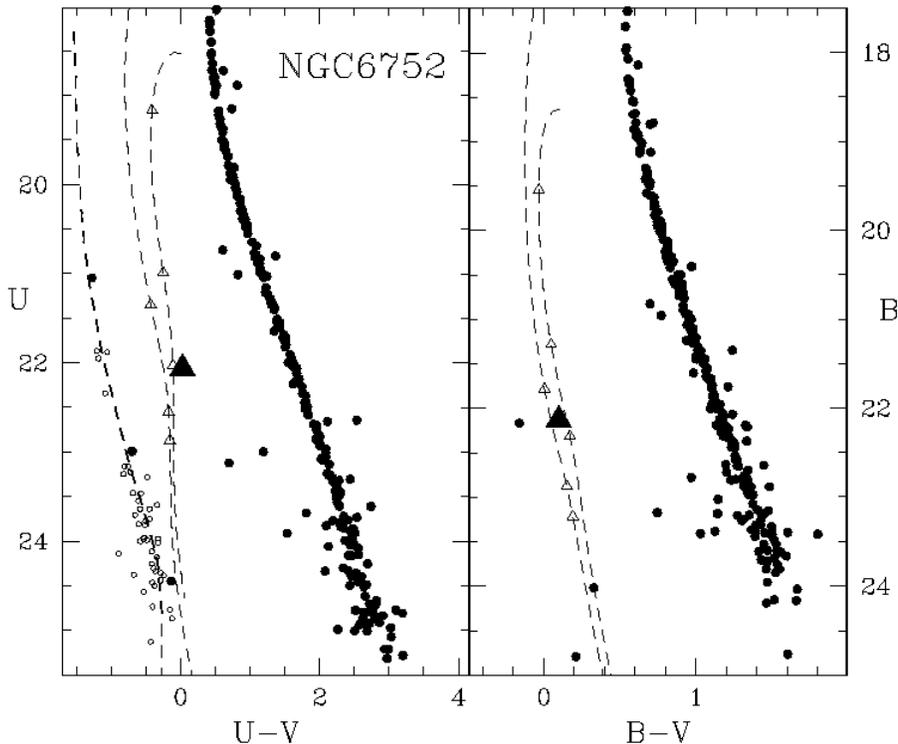}
\caption{$(U,U-V)$ and $(B,B-V)$ CMDs 
  for the stars identified in a region
of $80'' \times 80'' $ centered at the nominal position of the
PSR J1911-5958A in NGC6752. The optical counterpart to the pulsar companion
(COM J1911-5958A) is marked with a {\it large  filled
triangles}.  The {\it heavy dashed line} is the CO-WD cooling 
sequence from  Wood (1995);
the two {\it light dashed lines} are the cooling tracks for He WD
masses 0.197 and 0.172 $M_{\odot}$ (Serenelli et al. 2002).
(From Ferraro et al 2003b).}   
 \end{figure}

\begin{figure}[!ht]
  \plotone{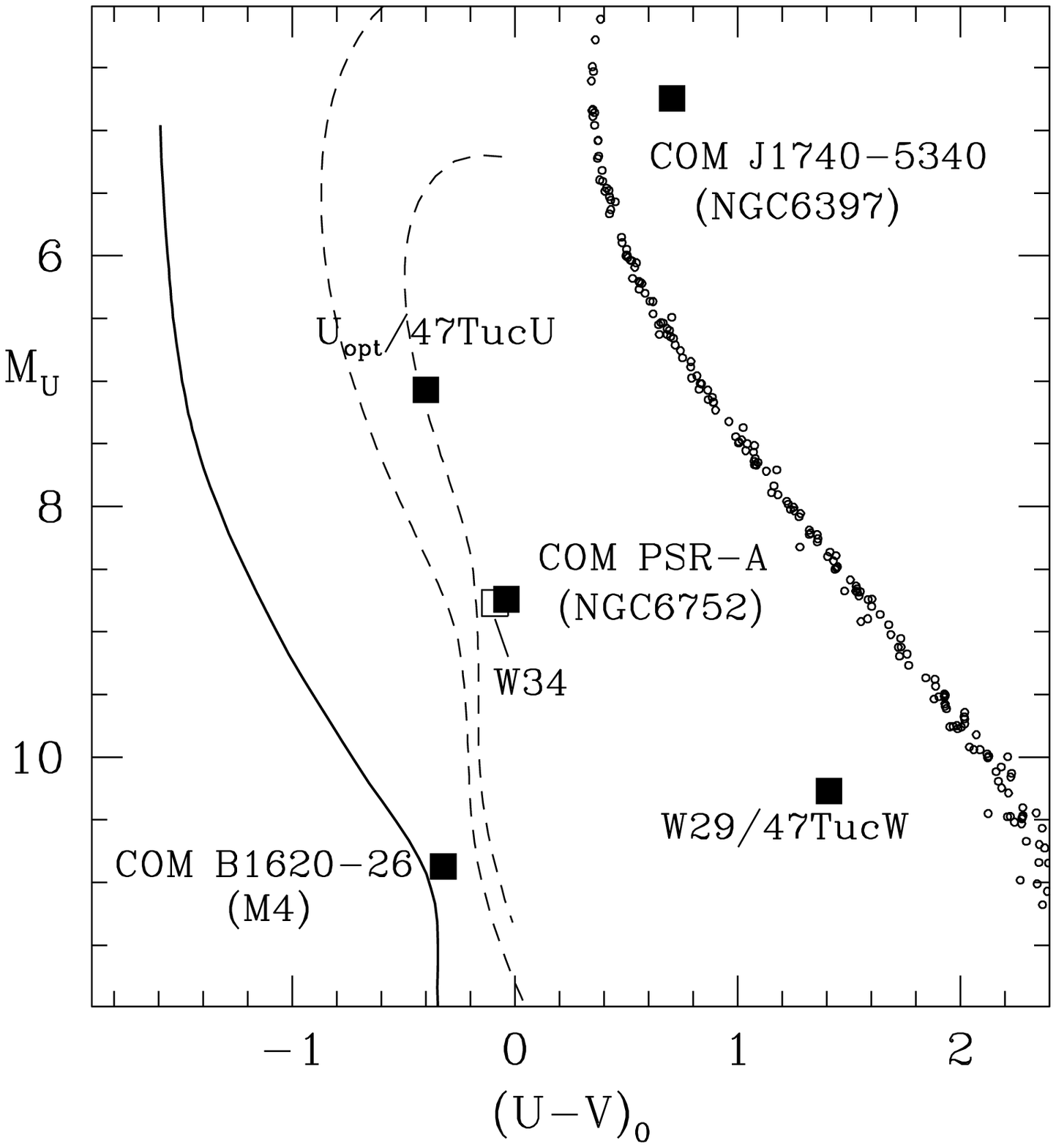}
  \caption{ All the optical counterparts to
  MSP companion detected so far in GCs are plotted (as 
  large filled squares) in the $(M_U,U-V)$ plane. The 
  cooling sequences for He-WD (from Serenelli et
al. 2002) and  for the  CO-WD  (from Wood 1995) are also 
replotted.
 MS stars of NGC6752 are  plotted  
for reference. (From Ferraro et al 2003b).}
\end{figure}

By using deep high-resolution multiband images taken at the
ESO VLT we recently identified 
the optical binary companion to
 PSR-A (COM-PSR-A, Ferraro et al 2003c). 
The object  turns out to be a blue
star whose position in the CMD is consistent with
the cooling sequence of a low mass ($M\sim 0.17-0.20M_{\odot}$), low
metallicity Helium WD (He-WD) 
(see Figure 23, see also Bassa et al 2003).
 The anomalous position of  PSR-A
with respect to the GC center 
suggested that this system has been  recently ($< 1$ Gyr)  ejected
from the cluster core as the result of a strong dynamical
interaction. 

\subsection{Photometric properties of MSP companion in GCs} 
 
The companion to 
PSR-A
is the second He-WD which has been found to orbit a  MSP
 in GCs.  Curiously, both objects  lie on the
same mass He-WD cooling sequence.   

Since the first discovery of $U_{opt}$, 
in 47 Tuc (Edmonds et al 2001), the zoo 
of the optical MSP counterparts in
GCs is rapidly enlarging. 
 Figure 24 shows a comparison of the photometric
properties of the available optical identifications of MSP companions
hosted in GCs. Note that we also include  an additional potential MSP
companion (W34 in 47 Tuc) discussed by
Edmonds et al. (2003).  2 out of the 5 sources are really peculiar:
the bright object in NGC6397 (which is as luminous as the turn
off stars and shows quite red colors) and the faint W29 in 47 Tuc,
which is also too red to be a He-WD (Edmonds et al 2002).  
 Indeed, $U_{opt}$ and  COM-PSR-A  
lie nearly on the same mass He-WD cooling sequence and W34 in 47 Tuc
curiously shares the same photometric properties of  COM-PSR-A. Indeed,
if confirmed as a MSP companion, W34 would be the third He-WD
companion orbiting a MSP in GCs roughly located on the same-mass
cooling sequence.  If further supported by additional cases, this
evidence could confirm that a low mass $\sim
0.15-0.2~M_{\odot}$ He-WD
orbiting a MSP is the favored system generated by the recycling
process of MSPs  
in GCs (Rasio et al 2000).

\subsection{MSPs as probes of the cluster dynamics}  

Among the list of the  GCs harboring MSP, NGC6752
presents a number of surprising features:

\begin{itemize}

\item  NGC~6752 hosts 5 known MSP (hereafter PSR A, B, C, D,
E, D'Amico et al. 2001; D'Amico et al. 2002).

\item As discussed in the previous section, NGC6752 harbors the two most 
 off-centered pulsars (PSR-A and PSR-C) 
 among the sample of 
MSPs whose position in the corresponding cluster is known.

\item   In the plane of the sky,  the 
positions  of the 
other 3 known MSPs in the cluster (PSR-B, D, E, 
all isolated pulsars, see D'Amico et
al. 2001a, 2002)  are close to
the cluster center, as expected from  mass
segregation in the cluster. D'Amico et al. (2002)
found two of them
(PSR-B and E) showing  large {\it negative} values of $\dot{P}$, implying
that the pulsars are experiencing an acceleration with a line-of-sight
component $a_l$ directed toward the observer and a magnitude
significantly larger than the positive component of $\dot{P}$ due to
the intrinsic pulsar spin-down.

\item By combining HST and wide field 
observations, Ferraro et al (2003d)
constructed the
most extended ($0'-27'$) and complete radial density profile ever
obtained for this cluster (see
Figure 25). 
The observed radial density profile   shows a
 significant deviation 
  from a canonical King model in the innermost region,
indicating, beyond any doubt, that the cluster core  is
experiencing the collapse phase.  On the basis of this data set 
Ferraro et al (2003d)
 also
re-determined the center of gravity $C_{\rm grav}$, which 
turns out to be $\sim
10\arcsec$ S and $\sim 2\arcsec$ E of the $C_{\rm lum}$ reported by
Djorgovski (1993).  Interestingly, the barycenter of
the 9 innermost X-ray sources detected by {\it Chandra} (Pooley et al
2002) is located only $1\farcs 9$ from the new $C_{\rm grav}.$
 
\end{itemize}

\begin{figure}[!ht]
  \plotone{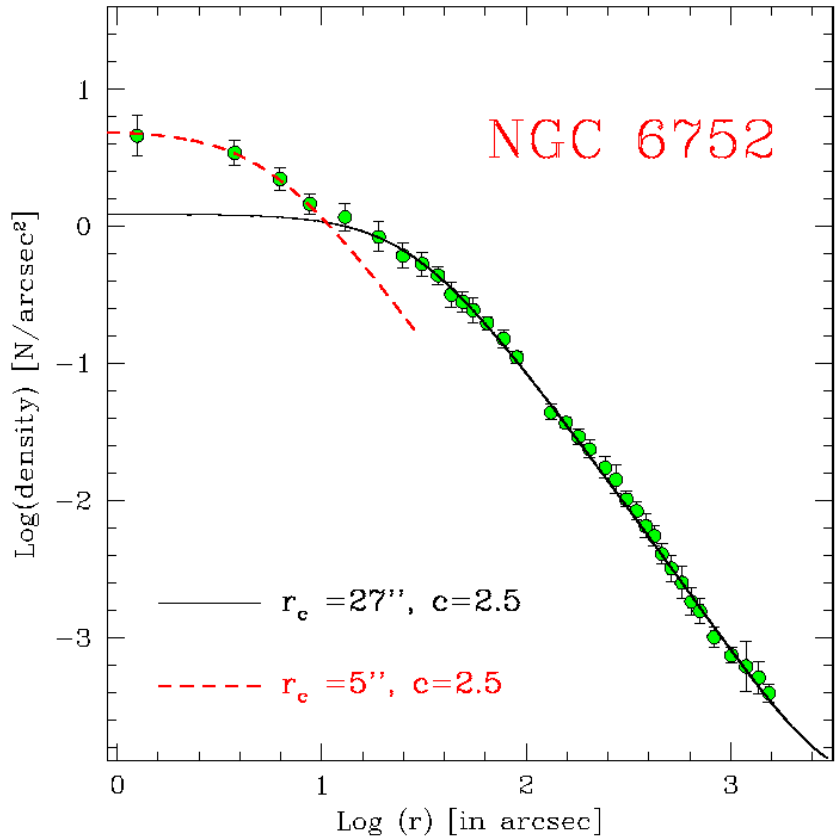}
  \caption{ Observed radial density profile with respect to the the
adopted $C_{\rm grav}$.  The solid line is the best fit King model
($r_c=28\arcsec$ and $c=1.9$) to the outer points (that is, for
distance larger than $10\arcsec$ from the cluster center).  The King
model has been combined with a constant star density of $ 0.25~{\rm
stars\,arcmin^{-2}}$ in order to account for the flattening of the
density profile in the extreme outer region ($r>16\farcm5$) of the
cluster. The dashed line is the best fit King model
to the 4 innermost points. (From Ferraro et al 2003d).}
\end{figure}

\begin{figure}[!ht]
  \plotone{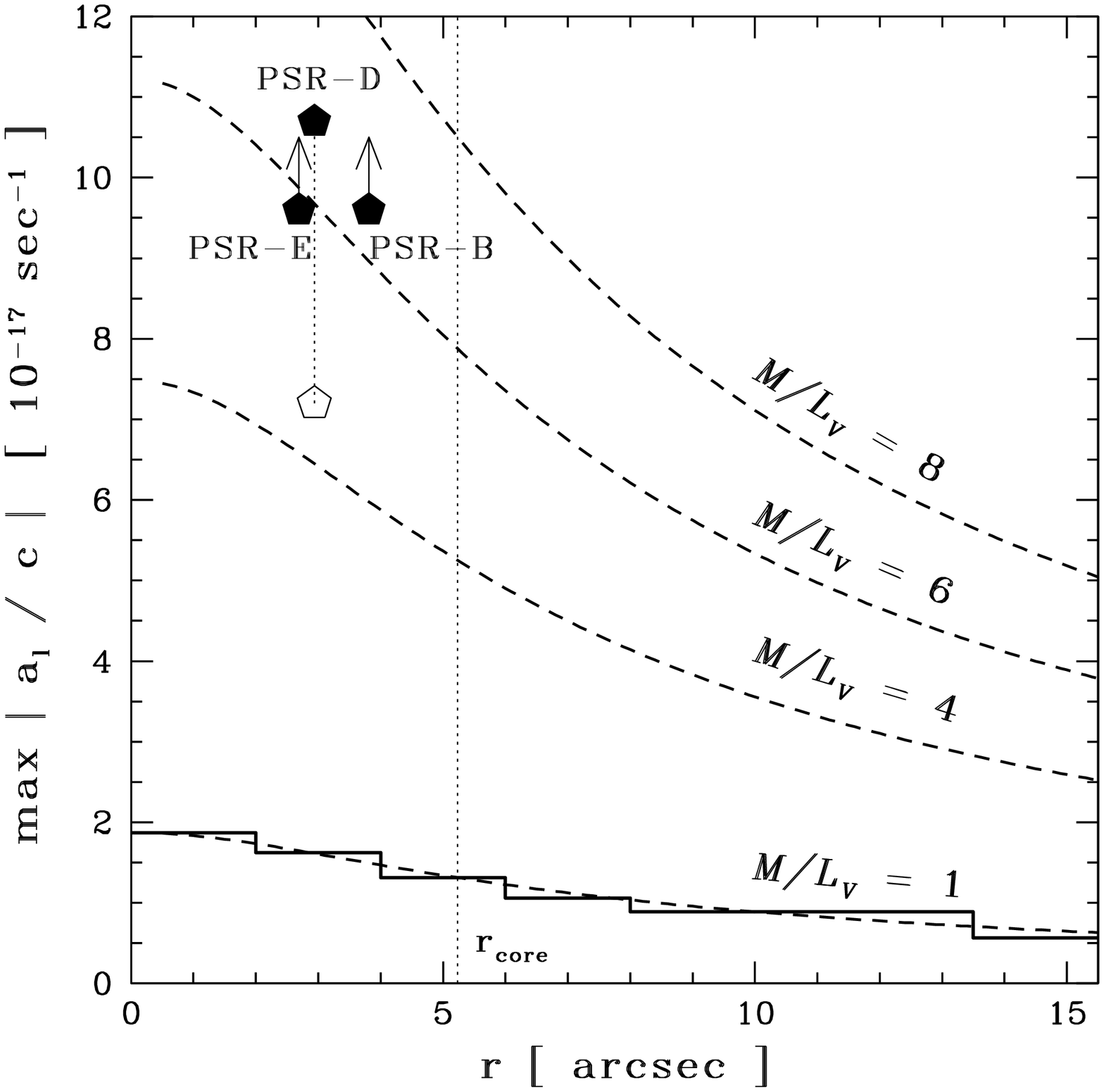}
  \caption{ Maximum line-of-sight acceleration $|a_{{l}_{max}}/c| =
|\dot{P}/P|$ versus radial offset with respect to the center of
NGC~6752. The histogram represents the prediction based on the star
density profile  (normalized to the central surface
brightness) assuming a unity mass-to-light ratio.  The
dashed lines are analytical fits to the optical observations, labeled
according to the adopted mass-to-light ratio.  The measured values of
$\dot{P}/P$ (filled pentagons, D'Amico et al. 2002) in the two MSPs
with negative $\dot{P}$ (PSR-B and E) can be reproduced only for
${\cal{M}}/{\cal L}_V> 6-7.$ The open pentagons show our best
guessed range of maximum $|a_l/c|$ for PSR-D: the upper value is
calculated assuming a negligible intrinsic positive $\dot P_{sd}$; the
lower value is estimated taking into account intrinsic $\dot P_{sd}$
from the observed scalings between X-ray luminosity and spin-down power
for MSPs (see D'Amico et al. 2002 and reference therein).  
(from Ferraro et al
2003d).}
\end{figure}

On the basis of these new results,  Ferraro et al (2003d)
suggested 
two viable explanations of the observed negative $\dot{P}$: {\it (i)}
the accelerating effect of the cluster gravitational potential well or
{\it (ii)} the presence of some close perturbator(s) exerting a
gravitational pull on the pulsars.

\smallskip
{\it Case (i): Overall effect of the GC potential well } The
hypothesis that the line-of-sight acceleration of the MSPs with
negative $\dot{P}$ is dominated by the cluster gravitational potential
has been routinely applied to many globulars.  In the case of NGC6752,
under the hypothesis that the line-of-sight acceleration of PSR-B and
PSR-E are entirely due to the cluster gravitational potential, a
${\cal{M}}/{\cal L}_V {\sim} 6$--7 is inferred (see Figure
26). Collapsed globulars
typically show values of ${\cal{M}}/{\cal L}_V {\sim} 2$--3.5 (Pryor \&
Meylan 1993). Assuming such a value, the expected total mass located
within the inner $r_{\perp,B}=0.08$ pc of NGC~6752 (equivalent to the
projected displacement of PSR-B from $C_{\rm grav}$) would be $\sim
1200$--$2000\,{M_\odot}$. On the other hand, the observed
${\cal{M}}/{\cal{L}}_V \sim 6$--7 implies the existence of an
additional $\sim 1500$--$2000\,{M_\odot}$ of low-luminosity matter
segregated inside the projected radius $r_{\perp,B}.$ This extra
amount of mass cannot be a relatively massive ($> 10^3\,{M_\odot}$)
black hole (BH), since it would produce a central power-law cusp in the
radial density profile, which is not observed  (see
Figure  25). The high ${\cal{M}}/{\cal{L}}_V$ ratio could be 
more likely due
to a very high central concentration of dark remnants of stellar
evolution, like NS and heavy ($\sim 1.0\,{M_\odot}$)
WD, which sank into the NGC~6752 core during the
cluster dynamical evolution (as also proposed for M15 by Gerssen et
al. 2003 and Baumgardt et al. 2003).

\smallskip
{\it Case (ii): Local perturbator(s)} One can imagine as an
alternative
possibility that the acceleration imparted to PSR-B and PSR-E is due
to some local perturbators.  Could a {\it single} object,
significantly more massive than a typical star in the cluster
simultaneously produce the accelerations detected both in PSR-B and
PSR-E?  Recently, Colpi, Possenti \& Gualandris (2002) suggested the
presence of a binary BH of moderate mass ($M_{\rm
bh+bh}\sim 100$--$200\,{ M_\odot}$) in the center of NGC~6752, in
order to explain the unexpected position of PSR-A in the outskirts
of the cluster.  The projected separation of PSR-B and PSR-E is only
$d_{\perp}=0.03$ pc.  A binary BH of total mass $M_{\rm bh+bh}$,
approximately located in front of them within a distance of the same
order of $d_{\perp}$, could  accelerate both  pulsars without
leaving any observable signature on the cluster density
profile.
  As the BH binarity ensures a large cross section to
interaction with other stars, the recoil velocity $v_{\rm rec}$ due to
a recent dynamical encounter could explain the offset position (with
respect to $C_{\rm grav}$) of the BH. However this scenario
has allow probability to occur: 
indeed, placing the BH at random within the core
gives roughly a 1\% chance that it would land in the right location 
to   produce the observed pulsar accelerations.

\acknowledgements            
 %%% Text of acknowledgements runs on after this command.

 It is a pleasure to thank Bob Rood, Alison Sills,
Michela Mapelli, Monica Colpi,  
 Giacomo Beccari, Elena Sabbi,
Andrea Possenti, Nichi D'Amico and the many other friends
who are collaborating to this project. 
This research was supported  by the {\it Ministero della
Istruzione dell'Universit\`a e della Ricerca} (MIUR).
 A special thanks to Livia for her lovely support.

%%% THE BIBLIOGRAPHY
%%%
%%% CONSULT SECTION 3 OF   manual_cozumel2005.tex    FOR HOW TO USE NATBIB.
%%% AUTHORS ARE ENCOURAGED TO USE EITHER THE "THEBIBLIOGRAPY" ENVIRONMENT
%%% or THE BIBTEX ENVIRONMENT. 

\end{document}